\documentclass[twocolumn]{aastex63}

\graphicspath{{./}{figures/}}
\usepackage{color}
\usepackage{soul}
\usepackage{natbib}
\usepackage{hyperref}
\usepackage{amsmath}
\usepackage{xspace}
\usepackage{lineno}
\usepackage{graphicx}
\usepackage{enumitem}
\usepackage{amssymb}
\usepackage{xifthen}
\usepackage{hyperref}
\usepackage[normalem]{ulem}
\usepackage{comment}

\hypersetup{    
  colorlinks      = {true},
  linkcolor       = {blue},
  citecolor       = {blue},
  urlcolor        = {blue},
}

\newcommand{\code}[1]{\texttt{#1}\xspace}

\newcommand{\resub}[1]{{#1}}

\newcommand{\SSSSS}{${S}^5$\xspace}
\newcommand{\gaia}{\textit{Gaia}\xspace}

\newcommand{\unit}[1]{\ensuremath{\mathrm{\,#1}}\xspace}
\newcommand{\feh}{\unit{[Fe/H]}}
\newcommand{\teff}{\ensuremath{T_\mathrm{eff}}\xspace}
\newcommand{\logg}{\ensuremath{\log\,g}\xspace}

\newcommand{\kms}{\unit{km\,s^{-1}}}

\newcommand{\msun}{\unit{M_\odot}}
\newcommand{\lsun}{\unit{L_\odot}}

\shorttitle{One Dozen Streams in \SSSSS}
\shortauthors{T.~S.~Li et al.}

\begin{document}

\title{\SSSSS: The Orbital and Chemical Properties of One Dozen Stellar Streams}

\author[0000-0002-9110-6163]{Ting~S.~Li}
\affiliation{Department of Astronomy and Astrophysics, University of Toronto, 50 St. George Street, Toronto ON, M5S 3H4, Canada}
\affiliation{Observatories of the Carnegie Institution for Science, 813 Santa Barbara St., Pasadena, CA 91101, USA}
\affiliation{Department of Astrophysical Sciences, Princeton University, Princeton, NJ 08544, USA}
\affiliation{NHFP Einstein Fellow}
\author[0000-0002-4863-8842]{Alexander~P.~Ji}
\affiliation{Department of Astronomy \& Astrophysics, University of Chicago, 5640 S Ellis Avenue, Chicago, IL 60637, USA}
\affiliation{Kavli Institute for Cosmological Physics, University of Chicago, Chicago, IL 60637, USA}
\author[0000-0002-6021-8760]{Andrew~B.~Pace}
\affiliation{McWilliams Center for Cosmology, Carnegie Mellon University, 5000 Forbes Ave, Pittsburgh, PA 15213, USA}
\author[0000-0002-8448-5505]{Denis~Erkal}
\affiliation{Department of Physics, University of Surrey, Guildford GU2 7XH, UK}
\author[0000-0003-2644-135X]{Sergey~E.~Koposov}
\affiliation{Institute for Astronomy, University of Edinburgh, Royal Observatory, Blackford Hill, Edinburgh EH9 3HJ, UK}
\affiliation{Institute of Astronomy, University of Cambridge, Madingley Road, Cambridge CB3 0HA, UK}
\affiliation{Kavli Institute for Cosmology, University of Cambridge, Madingley Road, Cambridge CB3 0HA, UK}
\author[0000-0003-2497-091X]{Nora~Shipp}
\affiliation{MIT Kavli Institute for Astrophysics and Space Research, Cambridge, MA 02139, USA}
\affiliation{Department of Astronomy \& Astrophysics, University of Chicago, 5640 S Ellis Avenue, Chicago, IL 60637, USA}
\affiliation{Kavli Institute for Cosmological Physics, University of Chicago, Chicago, IL 60637, USA}
\author[0000-0001-7019-649X]{Gary~S.~Da~Costa}
\affiliation{Research School of Astronomy and Astrophysics, Australian National University, Canberra, ACT 2611, Australia}
\affiliation{Centre of Excellence for All-Sky Astrophysics in Three Dimensions (ASTRO 3D), Australia}
\author[0000-0001-8536-0547]{Lara~R.~Cullinane}
\affiliation{Research School of Astronomy and Astrophysics, Australian National University, Canberra, ACT 2611, Australia}
\author[0000-0003-0120-0808]{Kyler~Kuehn}
\affiliation{Lowell Observatory, 1400 W Mars Hill Rd, Flagstaff,  AZ 86001, USA}
\affiliation{Australian Astronomical Optics, Faculty of Science and Engineering, Macquarie University, Macquarie Park, NSW 2113, Australia}
\author[0000-0003-3081-9319]{Geraint~F.~Lewis}
\affiliation{Sydney Institute for Astronomy, School of Physics, A28, The University of Sydney, NSW 2006, Australia}
\author[0000-0002-6529-8093]{Dougal~Mackey}
\affiliation{Research School of Astronomy and Astrophysics, Australian National University, Canberra, ACT 2611, Australia}
\author[0000-0002-8165-2507]{Jeffrey~D.~Simpson}
\affiliation{School of Physics, UNSW, Sydney, NSW 2052, Australia}
\affiliation{Centre of Excellence for All-Sky Astrophysics in Three Dimensions (ASTRO 3D), Australia}
\author[0000-0003-1124-8477]{Daniel~B.~Zucker}
\affiliation{Department of Physics \& Astronomy, Macquarie University, Sydney, NSW 2109, Australia}
\affiliation{Macquarie University Research Centre for Astronomy, Astrophysics \& Astrophotonics, Sydney, NSW 2109, Australia}
\author[0000-0001-6957-1627]{Peter~S.~Ferguson}
\affiliation{Physics Department, University of Wisconsin-Madison, 1150 University Avenue Madison, WI 53706, USA}
\author[0000-0002-3430-4163]{Sarah~L.~Martell}
\affiliation{School of Physics, UNSW, Sydney, NSW 2052, Australia}
\affiliation{Centre of Excellence for All-Sky Astrophysics in Three Dimensions (ASTRO 3D), Australia}
\author[0000-0001-7516-4016]{Joss~Bland-Hawthorn}
\affiliation{Sydney Institute for Astronomy, School of Physics, A28, The University of Sydney, NSW 2006, Australia}
\affiliation{Centre of Excellence for All-Sky Astrophysics in Three Dimensions (ASTRO 3D), Australia}
\author[0000-0002-1322-3153]{Eduardo~Balbinot}
\affiliation{Kapteyn Instituut, Rijksunversiteit Groningen, Postbus 800, NL-9700AV Groningen, the Netherlands}
\author[0000-0001-6584-6144]{Kiyan~Tavangar}
\affiliation{Department of Astronomy \& Astrophysics, University of Chicago, 5640 S Ellis Avenue, Chicago, IL 60637, USA}
\affiliation{Kavli Institute for Cosmological Physics, University of Chicago, Chicago, IL 60637, USA}
\author[0000-0001-8251-933X]{Alex~Drlica-Wagner}
\affiliation{Fermi National Accelerator Laboratory, P.O.\ Box 500, Batavia, IL 60510, USA}
\affiliation{Department of Astronomy \& Astrophysics, University of Chicago, 5640 S Ellis Avenue, Chicago, IL 60637, USA}
\affiliation{Kavli Institute for Cosmological Physics, University of Chicago, Chicago, IL 60637, USA}
\author{Gayandhi~M.~De~Silva}
\affiliation{Australian Astronomical Optics, Faculty of Science and Engineering, Macquarie University, Macquarie Park, NSW 2113, Australia}
\affiliation{Centre of Excellence for All-Sky Astrophysics in Three Dimensions (ASTRO 3D), Australia}
\author[0000-0002-4733-4994]{Joshua~D.~Simon}
\affiliation{Observatories of the Carnegie Institution for Science, 813 Santa Barbara St., Pasadena, CA 91101, USA}

\collaboration{21}{(\SSSSS Collaboration)}

\correspondingauthor{T.~S.~Li}
\email{ting.li@astro.utoronto.ca}



\begin{abstract}
   
We report the kinematic, orbital, and chemical properties of 12 stellar streams with no evident progenitors, using line-of-sight velocities and metallicities from the Southern Stellar Stream Spectroscopic Survey (\SSSSS), proper motions from \gaia EDR3, and distances derived from distance tracers or the literature. This data set provides the largest homogeneously analyzed set of streams with full 6D kinematics and metallicities.
All streams have heliocentric distances between ${\sim}10-50$ kpc. 
The velocity and metallicity dispersions show that half of the stream progenitors were disrupted dwarf galaxies (DGs), while the other half originated from disrupted globular clusters (GCs), \resub{hereafter referred to as DG and GC streams}. Based on the mean metallicities of the streams and the mass-metallicity relation, the luminosities of the progenitors of the DG streams range between \resub{Carina and Ursa Major I} ($-9.5\lesssim M_V\lesssim-5.5$).
Four of the six GC streams have mean metallicities of $\feh < -2$, more metal-poor than typical Milky Way (MW) GCs at similar distances. 
Interestingly, the 300S and Jet GC streams are the only streams on retrograde orbits in our dozen stream sample.
Finally, we compare the orbital properties of the streams with known DGs and GCs in the MW, finding several possible associations. Some streams appear to have been accreted with the recently discovered Gaia-Enceladus-Sausage system, and others suggest that GCs were formed in and accreted together with the progenitors of DG streams whose stellar masses are similar to Draco to Carina ($\sim10^5-10^6\msun$).
\end{abstract}



\section{Introduction} \label{sec:intro}

The hierarchical model of galaxy formation posits that the Milky Way (MW) was assembled through the accretion and disruption of many smaller systems \citep{Peebles:1965,Press:1974,Searle:1978,Blumenthal:1984}, such as dwarf galaxies (DGs) and globular clusters (GCs). Systems disrupted relatively recently can manifest as stellar streams, which provide a snapshot of accretion that can be compared directly with theoretical models of structure formation \cite[e.g.,][]{Johnston:1998,Freeman2002ARA&A,Bullock:2005} and bridge the gap between the smooth stellar halo and the present-day MW satellite population. Furthermore, the spatial and kinematic properties of stellar streams are sensitive to the mass and three-dimensional shape of the MW's gravitational field \citep[e.g.,][]{Koposov:2010,Erkal:2016a,Bovy:2016,Bonaca2018}. Gaps and kinks in stellar streams can also reveal the existence of low-mass dark matter substructure throughout the halo \cite[e.g.,][]{Ibata2002,Johnston2002,Carlberg:2009,Varghese2011,Erkal:2016b}, a key prediction of the cold dark matter model. 

As a result of the various ground-based photometric surveys in the past two decades, as well as the recent \gaia data releases, the number of known MW stellar streams has increased substantially to over sixty \citep[e.g.,][]{Grillmair:2016, Shipp:2018, Mateu:2017, Ibata2019}.  Some of these streams clearly emanate from known DGs or GCs (e.g., the Sagittarius and Palomar 5 streams), but most streams lack an obvious progenitor. 
\resub{While imaging surveys can provide on-sky positions, distances and proper motions (PMs), spectroscopic observations can reveal line-of-sight velocities and chemical properties of the streams.}
However, spectroscopic observations of \resub{individual stars in} stellar streams are extremely challenging due to the \resub{low stellar density} and large angular extents (${>}\,10\degr$) of these systems, particularly for the more distant (and thus fainter) streams beyond a heliocentric distance of 10 kpc. 
Prior to 2018, only a handful of stellar streams had spectroscopic measurements:
Sagittarius \citep{Majewski:2003}, Palomar 5 \citep{Odenkirchen:2009, Kuzma:2015,Ibata:2016,Li:2017}, Cetus \citep{Newberg:2009,Koposov:2012,Yam:2013,Li:2017}, GD-1 \citep{Koposov:2010,Li:2017}, Orphan \citep{Newberg:2010,Casey2013, Casey2014, Li:2017}, Ophiuchus \citep{Sesar:2015}, and the $300~\kms$ stream \citep{Simon:2011,Frebel2013,Fu:2018}. 
Even for these spectroscopically confirmed streams, spectroscopic measurements are often quite sparse due to the challenges of efficiently targeting member stars. 
This has made it hard to answer some of the most basic of questions, such as whether these streams are originated from disrupting GCs or dissolving DGs.

The Southern Stellar Stream Spectroscopic Survey (\SSSSS, \citealt{Li2019}) was initiated in 2018 using the Two-degree Field (2dF) fiber positioner \citep{Lewis:2002} coupled with the dual-arm AAOmega spectrograph \citep{Sharp:2006} on the 3.9-m Anglo-Australian Telescope (AAT). This ongoing survey pursues a complete census of known streams in the Southern Hemisphere and has so far produced observations of more than 20 stellar streams. In this paper, we present the most recent measurements of the stream properties for twelve stellar streams that have a robust detection of spectroscopic members from \SSSSS: ATLAS-Aliqa Uma (AAU), Elqui, Indus, Jet, Jhelum, Orphan-Chenab, Ophiuchus, Palca, Phoenix, Turranburra,  Willka Yaku, and the $300~\kms$ stream (hereafter 300S). In addition, we also report the failure to detect spectroscopic stream members in the following stream candidates: Ravi, Wambelong, and Styx. Except for Ophiuchus at $\sim8$ kpc, all of the streams we present here are distant, with heliocentric distances between 10 and 50 kpc.
This is currently the largest homogeneously analyzed set of streams with full 6D kinematics and metallicities. Here we only focus on streams that have no obvious progenitors inside the streams, although \SSSSS has also observed several streams with known progenitors (i.e., GCs or DGs presenting tidal tails, but which are not fully disrupted; e.g., \citealt{Ji2021}).

In Section \ref{sec:data}, we describe the stream observables from the data, such as distance, velocity and metallicities. In Section \ref{sec:results}, we discuss the properties of the progenitors of these streams based on their kinematic and metallicity results from Section \ref{sec:data}.  In Section \ref{sec:discuss}, we discuss our findings based on the measurements, and we conclude in Section \ref{sec:conclusion}.

\section{Data in 6+1D}\label{sec:data}

\subsection{\SSSSS Data}

The data used in this paper are from an internal \SSSSS data release (iDR3.1) where the analysis has been improved compared to \citet[][previously iDR1.5]{Li2019};
\SSSSS observations taken between 2018 and 2020 are included. For details of the updates in the spectral fitting, we refer readers to \citet{Ji2021}.
In short, we simultaneously model spectra from both the red and blue arms of AAOmega as well as repeated observations of the same object from different nights with \code{rvspecfit} \citep{rvspecfit}, which provides radial velocities (RVs), effective temperatures ($\teff$), surface gravities ($\logg$), stellar metallicities ($\feh$), and alpha abundance ([$\alpha$/Fe]) estimates for each star.  The blue spectra make the metallicities from \code{rvspecfit} more accurate than for iDR1.5 (where metallicities were derived using only the spectra from the red arm). 
We also derived preliminary distance measurements for individual stars as the updates from iDR2.2 (used in \citealt{Ji2021}) to iDR3.1. The distances are determined using a full combination of parallax from \gaia, the spectra from \SSSSS, the available optical/infrared photometry for each star 
and MIST isochrones \citep{Choi2016}. We estimate a distance uncertainty of $\sim25\%$ for distant stars with poor parallax measurements from this spectrophotometric derivation and we are actively working to improve the distance uncertainty for future releases. 

We also determine calcium triplet (CaT) metallicities from equivalent widths (EWs) and the \citet{Carrera13} calibration for all the red giant branch stream member stars. CaT metallicities are more accurate than those from \code{rvspecfit} when the distances to the stars are known \citep{Li2019,Ji2020b}.
EWs were measured by fitting a Gaussian plus Lorentzian function \citep[e.g.,][]{eri2}. A systemic uncertainty floor of 0.19 \AA, computed from repeated observations, is added in addition to the covariance from the EW fit as the total uncertainty on the EW, and then transferred to the CaT metallicity uncertainty through error propagation using the \citet{Carrera13} calibration relation.
The absolute $V$ magnitudes are determined from \gaia EDR3 $G$, $Bp$, and $Rp$ photometry, first applying the color transformation function defined in \citet{Riello2021}, then dereddening using \citet{Schlegel:1998,Schlafly2011} and adding the distance moduli of individual stars detailed in Section \ref{sec:distance}.

\begin{figure*}[!htb]
    \centering
    \includegraphics[width=0.9\textwidth]{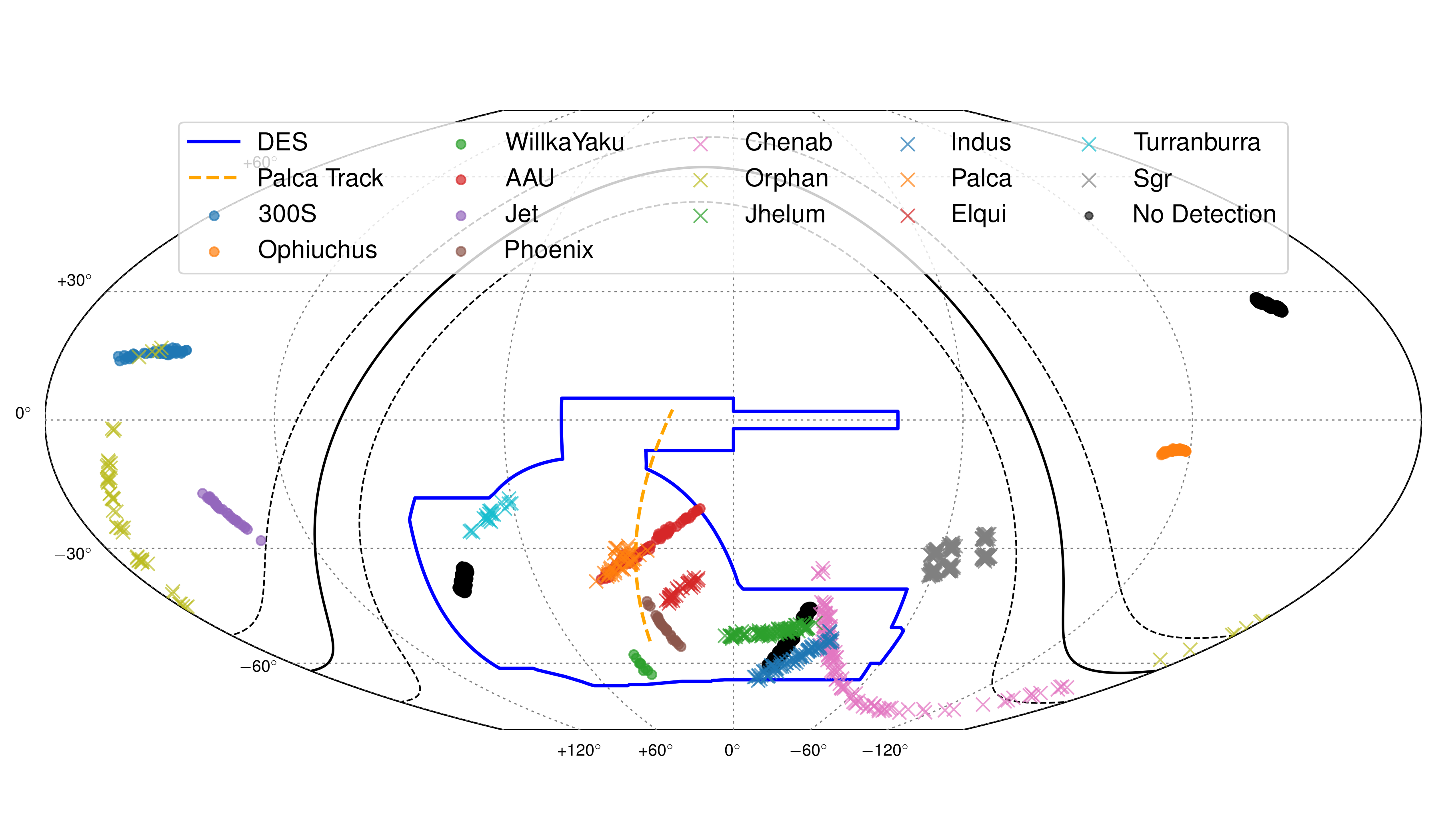}
    \caption{On-sky distribution of member stars identified with \SSSSS observations taken between 2018 and 2020 in a dozen stellar streams. Stars \resub{in streams that we consider to have had GC and DG progenitors} are shown as circles and crosses, respectively. Among these streams, nine are within the DES footprint (outlined in blue). Three streams, Wambelong, Ravi, and Styx, have no clear spectroscopic members identified; the coverage by \SSSSS so far on these streams is shown in the black \resub{filled circles}. Note that since no observations of Palca had been planned by \SSSSS, we have only identified Palca member stars in the AAU stream field (see \citep{Li2021} for more details). The Palca stream is much longer than the observed length, as indicated by the orange dashed line from \citet{Shipp:2018}.
    }
    \label{fig:members}
\end{figure*}

\subsection{Member Stars in 12 Streams}

We report the kinematic and chemical properties of the following one dozen streams that were observed in \SSSSS between 2018 and 2020: AAU, Elqui, Indus, Jet, Jhelum, Orphan-Chenab, Ophiuchus, Palca, Phoenix, Turranburra,  Willka Yaku, and 300S, most of which were first discovered in the Dark Energy Survey (DES, \citealt{Shipp:2018}), 
\resub{using a matched-filter in color-magnitude space derived from a synthetic isochrone of an old, metal-poor stellar population}.
The Sagittarius (Sgr) stream was also observed in \SSSSS, which we only show for comparison here given numerous previous studies on Sgr \citep[e.g.,][]{Hasselquist2019,CQYang2019,Johnson2020}.
\SSSSS had not completed planned observations of 300S, Turranburra, Jet, Palca, and Orphan-Chenab by the end of 2020, but enough data had been collected to allow member stars to be identified for this study.

Six of our streams have detailed membership studies published or in preparation.
For streams with published \SSSSS papers, we take the spectroscopic members directly from the references (\citealt{Li2021} for AAU and Palca, \citealt{Wan2020} for Phoenix). These studies use \gaia DR2, but switching to EDR3 makes no difference to their membership.
Detailed membership studies are underway by \SSSSS for Indus and Jhelum (A.~B.~Pace et al. in prep) and Elqui (S.~Martell et al. in prep), so we use stars with high membership probabilities ($P>0.8$) from those studies as stream members. \resub{The membership are computed using a probabilistic mixture model detailed in \citet{Wan2020}.} 
For Orphan-Chenab, we select member stars within $\pm15$ \kms and $\pm0.4$ mas~yr$^{-1}$ from the best RV and PM track determined in S.~E.~Koposov et al. (in prep) \resub{through fitting to RV data from \SSSSS and other spectroscopic surveys as well as fitting for stream PM in {\it Gaia} EDR3 data expanding the analysis from \citet{Koposov:2019} } . Since parts of the stream have velocities close to MW foreground stars, we also require a heliocentric \SSSSS iDR3.1 distance larger than 10 kpc to remove foreground main sequence stars. 
Note that Orphan and Chenab are a single stream \citep{Koposov:2019}, but since the stream is extremely long (over $200\degr$ on the sky), and the southern part is heavily perturbed by the Large Magellanic Clouds \citep[LMC,][]{Erkal2019}, we consider their properties separately to probe the possible kinematic and metallicity variation along the long stream. We define the Orphan and Chenab components of this stream as $\phi_1 > 0$ and $\phi_1 < 0$, respectively, with $\phi_1$ as defined in \citet{Koposov:2019}. 

For the remaining streams -- i.e., Jet, Turranburra, 300S, Ophiuchus, and Willka Yaku -- no comprehensive membership modeling has yet been performed. We select member stars with simple kinematic cuts using RVs from \SSSSS and PMs from \gaia EDR3, in a similar method as used for AAU in \citet{Li2021} and Phoenix in \citet{Wan2020}, to ensure a sample that is as pure as possible. With the exception of Turranburra, all other likely stream members are selected only kinematically (so that no metallicity bias is introduced). Turranburra's kinematics are almost identical to foreground disk stars, so we select member stars with a metallicity criterion of $\feh < -1.5$ to remove foreground stars. However, we believe that Turranburra is a real and distinct structure, given the low metallicities ($\feh < -2$) of most member stars.
The member selection here is not the most comprehensive, but rather a straightforward and simple choice to 1) confirm that the stream is a real signature in kinematic and metallicity space, and 2) identify a list of high purity (at a cost of a lower completeness) members to study the kinematic, orbital, and metallicity properties of the stream for further discussion in later sections. 

In Figure \ref{fig:members}, we show the on-sky location of the spectroscopic members in the dozen streams that are used in this work. To ensure the quality of the data, we only include member stars with signal-to-noise ratio (S/N) $> 5$ in red arm spectra in this work, though \SSSSS does find more member stars at lower S/N. 
\resub{For individual stars, the typical uncertainties range from $\sim1\kms$ at S/N $> 20$ to $\sim4\kms$ at S/N $\sim 5$ for RV, and from 0.15 dex at S/N $> 20$ to 0.4 dex at S/N $\sim 5$ for metallicites. We refer readers to \citet{Li2019} for reduction and validation of \SSSSS data.}
We also note that, as the observations in several streams have not yet finished, more spectroscopic members are expected in future \SSSSS releases and publications. 

\begin{deluxetable*}{l r r r r r r c c c c r}[!htb]
\tablecolumns{11}
\tablewidth{0pt}
\tabletypesize{\scriptsize}
\tablecaption{Kinematic and metallicity properties of one dozen streams observed with \SSSSS. \resub{As we probe the Orphan and Chenab components of Orphan-Chenab stream separately, there are thirteen rows for twelve streams presented here.} From left to right, the columns are the name of the stream, the distance relation adopted in this paper, average heliocentric distance ${\bar{d}_\odot}$, stream width $\sigma_w$, number of spectroscopic members in each stream, velocity dispersion $\sigma_\mathrm{vel}$, number of red-giant branch members that have CaT metallicities, mean CaT metallicities $\overline{\feh}_\mathrm{CaT}$, CaT metallicity dispersion $\sigma_{\mathrm{\feh}}$, 95th percentile of the CaT metallicity dispersion, whether the progenitor is a dwarf galaxy (DG) or globular cluster (GC), and analogs to the known MW satellite galaxies based on their mean metallicities (if applicable), respectively.} \label{table:stream_progenitor}
\tablehead{\\
Stream  &  Distance$^{(a)}$  & ${\bar{d}_\odot}$ & $\sigma_w^{(b)}$ & $N_\mathrm{mem}$  & $\sigma_\mathrm{vel}^{(c)}$   & $N_\mathrm{RGB}$  & $\overline{\feh}_\mathrm{CaT}$  &  $\sigma_{\mathrm{\feh}}$\  & $\sigma_{\mathrm{\feh}}$ & Progenitor & Progenitor \\
             &            &        (kpc)      &   (pc)  &    & (\kms)  &  &     &  & $(<95\%)$ & Type  &  Analog
}
\startdata
        300S  &           (d/1kpc)=$48.9952-0.2083\alpha$  &   17.2  &   110  &   53   &   $ 2.5_{-0.3}^{+0.4}$ &     52 & $-1.26_{-0.03}^{+0.03}$ & $0.04_{-0.02}^{+0.04}$ & 0.11 &     GC  &                 \\
   Ophiuchus  &                      $\mu=14.58-0.2(l-5)$  &    7.9  &     8  &  118   &   $ 2.4_{-0.3}^{+0.3}$ &     37 & $-1.80_{-0.03}^{+0.03}$ & $0.03_{-0.01}^{+0.03}$ & 0.09 &     GC  &                 \\
  WillkaYaku  &                                $\mu=17.8$  &   36.3  &   127  &    9   &   $ 0.4_{-0.4}^{+0.8}$ &      7 & $-2.05_{-0.07}^{+0.07}$ & $0.04_{-0.02}^{+0.07}$ & 0.18 &     GC  &                 \\
         AAU  &                   $\mu=16.67-0.034\phi_1$  &   23.8  &    96  &   85   &   $ 4.3_{-0.4}^{+0.4}$ &     66 & $-2.22_{-0.02}^{+0.02}$ & $0.05_{-0.03}^{+0.05}$ & 0.13 &     GC  &                 \\
         Jet  &                   $\mu=17.45-0.014\phi_1$  &   30.4  &    90  &   32   &   $ 0.7_{-0.5}^{+0.4}$ &     29 & $-2.38_{-0.03}^{+0.03}$ & $0.04_{-0.02}^{+0.05}$ & 0.12 &     GC  &                 \\
     Phoenix  &                   $\mu=16.26+0.008\phi_1$  &   17.9  &    53  &   26   &   $ 2.5_{-0.7}^{+0.7}$ &     20 & $-2.62_{-0.05}^{+0.05}$ & $0.03_{-0.02}^{+0.05}$ & 0.13 &     GC  &                 \\
      Orphan  &        track from Koposov et al. &   17.1  &   747  &   58   &   $ 4.1_{-0.5}^{+0.5}$ &     49 & $-1.85_{-0.07}^{+0.07}$ & $0.42_{-0.06}^{+0.07}$ & 0.53 &     DG  &          Leo II \\
      Chenab  &         track from Koposov et al.  &   32.8  &   493  &  125   &   $ 4.5_{-0.3}^{+0.5}$ &    109 & $-1.78_{-0.04}^{+0.04}$ & $0.28_{-0.03}^{+0.03}$ & 0.34 &     DG  &          Leo II \\
      Jhelum  &                                $\mu=15.4$  &   12.0  &   267  &   95   &   $13.7_{-1.1}^{+1.2}$ &     39 & $-1.83_{-0.05}^{+0.05}$ & $0.25_{-0.04}^{+0.05}$ & 0.34 &     DG  &          Leo II \\
       Indus  &                   $\mu=15.90-0.016\phi_1$  &   15.2  &   240  &   75   &   $ 7.6_{-0.6}^{+0.7}$ &     66 & $-1.96_{-0.05}^{+0.05}$ & $0.33_{-0.04}^{+0.05}$ & 0.41 &     DG  &           Draco \\
       Palca  &                                $\mu=17.8$  &   36.3  &  $\gtrsim1000$  &   42   &   $13.4_{-1.4}^{+1.9}$ &     32 & $-2.02_{-0.04}^{+0.04}$ & $0.13_{-0.07}^{+0.06}$ & 0.23 &     DG  &           Draco \\
       Elqui  &                   $\mu=18.48-0.043\phi_1$  &   51.0  &   472  &   34   &   $16.2_{-2.1}^{+2.3}$ &     33 & $-2.22_{-0.06}^{+0.06}$ & $0.27_{-0.05}^{+0.06}$ & 0.37 &     DG  &    Ursa Major I \\
 Turranburra  &                                $\mu=17.1$  &   26.3  &   288  &   22   &   $19.7_{-3.0}^{+3.9}$ &     15 & $-2.18_{-0.14}^{+0.13}$ & $0.39_{-0.09}^{+0.12}$ & 0.62 &     DG  &    Ursa Major I \\
\enddata
{\footnotesize \tablecomments{ 
$^{(a)}$ Distance relation derived in this work with BHB and RRL members (see text for details) except for 300S \citep[][$\alpha$ is the RA of each star in degrees]{Fu2019}, Ophiuchus \citep[][$l$ is the Galactic longitude of each star in degrees.]{Sesar:2015}, and Jet \citep{Ferguson2021}. For Orphan-Chenab, we adopt the distance relation from RRLs in Koposov et al. (in prep)
For the Palca stream, the distance modulus is computed based on members at ($\alpha$,$\delta$) ={($33\degr, -33\degr$)}. Note that the distance for other parts of the stream may be different.\\
$^{(b)}$ Stream width, defined as one standard deviation of a Gaussian profile, taken from different literature, 300S \citep{Fu2019}, Ophiuchus \citep{Bernard:2014}, Orphan \citep{Grillmair:2006}, and the rest from \citet{Shipp:2018}. \citet{Shipp:2018} did not measure the width of Palca stream. We therefore set it to be $\gtrsim 1000$ pc based on the measurements on Cetus stream from \citet{Yam:2013}.\\
$^{(c)}$ The velocity dispersion of Chenab and Orphan might be underestimated, as we selected member stars within $\pm$15 \kms from the RV track defined in Koposov et al. (in prep) to minimize the background contamination. We refer to Koposov et al. (in prep) for a more rigorous calculation.
}}
\end{deluxetable*}

\subsection{Distance}\label{sec:distance}

Distance is a crucial input for both the kinematic and chemical studies of these streams. 
Although \SSSSS iDR3.1 provides distances to individual stars, the uncertainties on individual stars are quite large (up to 25\%).
Therefore, in this work, we compute the stream distance measurements for all the streams that have DES photometry, using individual blue horizontal branch stars (BHBs) and RR Lyrae stars (RRLs), as described in \citet{Li2021}.

BHB members are identified as spectroscopic stream members with $g-r < 0$ that are not in the \gaia RRL catalogs \citep{Clementini2018,Holl2018}.
The distance modulus of each BHB is calculated using the relation from \citet{Belokurov:2016} and the dereddened DES DR2 photometry. 
We find RRL members by cross-matching spectroscopic members with the \gaia RRL catalogs.\footnote{Our RRL sample might be incomplete, as the RRLs could have very large velocity variability and therefore might not be classified as spectroscopic members.}
We then determine the distance moduli of the RRL stars using the relation from  \citet{Muraveva:2018}:
$$M_G = 0.32 \feh + 1.11$$
and the dereddened \gaia $G$-band magnitudes, using the color-dependent extinction corrections from \citet{Babusiaux2018} and $E(B-V)$ values from \citet{Schlegel:1998} and \citet{Schlafly2011}.
Since the metallicities derived from CaT EW require a distance input, we adopt a preliminary mean metallicity derived from BHB distances for the RRL distance derivation. We note that a 0.1 dex shift in metallicity will only impact the distance moduli by 0.03 mag for RRLs and is much smaller than the intrinsic dispersion of 0.17 mag found by \citet{Muraveva:2018}.

\resub{Among all the streams we have computed the distances with BHBs and RRLs, AAU has the most distance tracers, with 13 BHBs and 5 RRLs \citep{Li2019}. Other streams have typical several to a few ($<10$) BHBs and RRLs in combined.} We note that we only select BHB and RRL members from the spectroscopic sample. Therefore, more BHB and RRL members are likely present outside the AAT fields, especially for the streams with large stream widths, but we limit our selection to the stars that have line-of-sight velocities for a purer sample. 

We then transfer all the BHB and RRL members into stream coordinates with the transformations defined in \citet{Shipp2019}. If a distance gradient is obvious along the stream longitude $\phi_1$, we fit a first-order polynomial in $\phi_1$, which is useful for the orbital properties discussed later.
Otherwise, we use a single average distance for the entire stream. The latter could be either due to a lack of distance gradient along the stream or a lack of distance tracers in the spectroscopic sample. 
We then assign a distance to each stream member based on the adopted relation.
The adopted heliocentric distance relations and the average heliocentric distances to each stream are presented in Table \ref{table:stream_progenitor}. 
We note that all the updated distance relations here have a mean distance modulus no more than 0.1 mag different from \citet{Shipp:2018}, who computed the distance from isochrone fitting. 

DES photometry does not have full coverage for the following streams: Orphan-Chenab, 300S, Ophiuchus, Jet, and Sgr;
we thus adopt the distances from a variety of references.
For Orphan-Chenab, we take the distance track from Koposov et al. (in prep). The distances for 300S, Jet, and Ophiuchus are taken from the references listed in Table \ref{table:stream_progenitor}. For Sgr, we use the median distance from \SSSSS iDR3.1 for each Sgr field. Since there are a total of over 1000 member stars in Sgr stream fields, the median distance from \SSSSS iDR3.1 is sufficient to obtain Sgr's orbital properties.

\subsection{Streams with No Detection of Spectroscopic Members}

We failed to identify any clear member stars in the following three streams: Ravi, Wambelong, and Styx. Neither the Ravi nor Wambelong streams had conclusive PM signatures in \citet{Shipp2019} with \gaia DR2, and we cannot detect them in \gaia EDR3 with spectroscopic information either.
Interestingly, the stream fields for both Ravi and Styx show groups of stars with small velocity dispersions that are known to belong to other structures. Specifically, member stars of the Tucana~II DG are detected in the Ravi fields (at a much larger distance than Ravi's reported distance), and a group of B\"ootes III stars are detected in the Styx fields. These additional structures may have caused difficulties in finding member stars in these streams. Furthermore, the incomplete mapping of these streams may result in a failure to obtain spectroscopic confirmation. Only four out of seven Wambelong fields have been observed so far by \SSSSS; Styx is a stream in the Northern Hemisphere and therefore we have only obtained 3 fields that are accessible from the AAT; and Ravi's stream fields largely overlap with those of the Jhelum and Indus streams. In Figure \ref{fig:members}, we show the areas that have been covered by \SSSSS along these three streams as black shaded regions.
Deeper data and/or larger surveyed areas are needed to confirm the reality of these streams; alternatively, these streams might not be real structures.

\section{Results}\label{sec:results}

\subsection{Velocities and Metallicities}

After selecting all stream member stars, we use the RA, Dec, RVs, and \feh from \SSSSS for the individual member stars to determine the velocity dispersion, mean metallicity, and metallicity dispersion of each stream. Rather than fitting a comprehensive background model (which is carefully done in other \SSSSS papers), here we just take the dispersions from a high-purity sample of member stars directly.

For the velocity dispersion, we first fit the stream RVs as a function of $\phi_1$ with a second-order polynomial and subtract that off as the systemic velocity of the stream along $\phi_1$. Then we model the residual RV with a Gaussian distribution, including Gaussian velocity uncertainties on individual stars, to derive the velocity dispersion. The posterior on the velocity dispersion was derived with the Markov Chain Monte Carlo (MCMC) sampler \code{emcee}\citep{emcee}, similar to what has been done in kinematic studies of DGs \citep[e.g.,][]{eri2}.
We use a uniform prior in log-space for the velocity dispersion with a prior range of (0.01-100)\,\kms. The velocity dispersion of each stream is reported in Table \ref{table:stream_progenitor}.

Similarly, we derive the mean metallicity and metallicity dispersion of each stream assuming a Gaussian metallicity distribution function (independent of $\phi_1$) with both the \code{rvspecfit} and CaT metallicities. 
For most of the streams, the mean metallicities from the two different methods (i.e. \code{rvspecfit} and CaT) give consistent results within 0.2 dex. 
The CaT metallicities are usually more accurate, though they rely on an accurate distance determination. As an example, the mean heliocentric distance of the 300S stream members is $\sim$17 kpc from \citet{Fu2019}; if we increased the distance by $30\%$ to 22 kpc, then the mean metallicity of 300S would change from $\feh = -1.27$ (the mean value from the CaT measurements) to $\feh = -1.4$ (mean metallicity from \code{rvspecfit}). 

Given the established membership of stars, and thus well-known distances, we consider the metallicities and metallicity dispersions from the CaT to be more accurate than those from \code{rvspecfit}. We therefore report the mean metallicities and metallicity dispersions of the dozen streams from the CaT metallicities in Table~\ref{table:stream_progenitor}. Our metallicity discussion in Section \ref{sec:progenitor} is also based on these quantities. Several streams have unresolved CaT metallicity dispersions, so we also give the 95\% confidence upper limit on the metallicity dispersion in Table~\ref{table:stream_progenitor}.

Finally, we note that the velocity dispersion of Orphan-Chenab might be biased low as $|\Delta\mathrm{RV}|<15$\kms is used for the membership selection. Similarly, the mean metallicity and dispersion of Turranburra might be biased low, since we discarded any possible member stars with $\feh > -1.5$.

\subsection{Stream Progenitors}\label{sec:progenitor}

\begin{figure*}[!htb]
    \centering
    \includegraphics[width=1.0\textwidth]{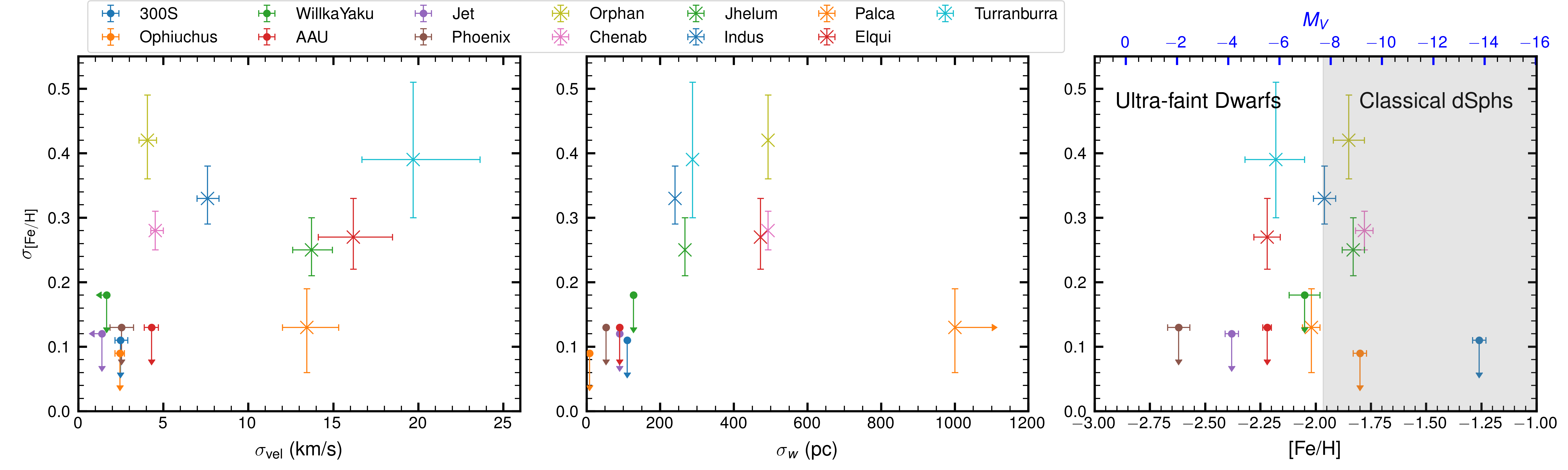}
    \caption{(left) Velocity dispersion \resub{($\sigma_\mathrm{vel}$)} versus metallicity dispersion \resub{($\sigma_\mathrm{\feh}$)} of one dozen streams. Six streams have both small velocity dispersions and metallicity dispersions and are classified as GC streams (circles); the other streams have larger metallicity dispersions and / or larger velocity dispersions and are classified as DG streams (crosses). If the dispersion is resolved, the median value is used for the symbol and the errors are taken from the 16th and 84th percentile. If the dispersion is not resolved, then the 95th percentile is used as the upper limit. (middle) Stream width \resub{($\sigma_w$)} versus metallicity dispersion of one dozen streams. GC stream have smaller stream width than the DG streams. (right) Mean CaT metallicity \resub{(\feh)} versus metallicity dispersion for these streams. The top axis shows the corresponding luminosity for progenitors of the DG streams, based on the empirical luminosity-metallicity relation \resub{for intact DGs} \citep{Norris2010, Kirby:2013}. $M_V = -7.7$ is used to separate classical dwarf spheroidals (dSphs) and ultra-faint dwarf (UFD) galaxies \citep{Simon:2019}. }
    \label{fig:feh}
\end{figure*}

As shown in Table \ref{table:stream_progenitor}, of the 12 streams presented here, half show an unresolved CaT metallicity dispersion (i.e., the dispersion is consistent with zero within the 2-sigma uncertainty). Furthermore, these streams have metallicity dispersions below 0.2 dex at the 95\% confidence level. We therefore conclude that the progenitors of these streams -- i.e., 300S, AAU, Jet, Ophiuchus, Phoenix, and Willka Yaku -- are likely to be GCs \resub{(hereafter referred to as GC streams)}. Most of these streams also show velocity dispersions below 3\,\kms. AAU has a higher dispersion, likely as a result of heating via perturbations \citep{Li2021}. These six streams are all clustered in the lower-left corner in the left panel of Figure \ref{fig:feh}. We use circle symbols to represent streams (or stream members) whose progenitors are GCs throughout this paper.

The remaining six streams -- Orphan-Chenab, Elqui, Indus, Palca, Turranbura, and Jhelum -- show resolved metallicity dispersions, so their progenitors are likely to be DGs \resub{(hereafter referred to as DG streams)}. These streams also show larger velocity dispersions than the GC streams in general. In particular, four streams have unusually high velocity dispersions of over 10 \kms, and are discussed in more detail in Section \ref{sec:largedispersion}. We use cross symbols to represent DG streams (or stream members) throughout this paper. We note that, since the Palca stream was not targeted on purpose, the Palca members were selected with an isochrone selection at the distance of AAU ($\Delta$DM = 1.1 closer than Palca). However, with a wide selection window in color and magnitude, most of the Palca members should be selected with the selection criteria for AAU. The brightest metal-rich members might be outside our selection window, which could result in a slightly lower metallicity dispersion.

Although the progenitor classification is primarily based on the metallicity dispersion, we see that both velocity dispersion and stream width also provide information about the progenitors. In the middle panel of Figure \ref{fig:feh}, we show the stream width\footnote{Here we define the stream width, $\sigma_w$, as one standard deviation of a Gaussian profile. Some other analyses instead quote the full width at half maximum (FWHM) $= 2.355 \sigma_w$.} of the dozen streams, compiled from the literature. All of the GC streams have a stream width $\sigma_w < 150$ pc, while the DG streams have $\sigma_w > 200$ pc.

In the right panel of Figure \ref{fig:feh}, we show the mean metallicities of these streams.  While the mean metallicities of the GC streams range from $-2.7$ to $-1.2$, DG streams in our sample show only a 0.5 dex spread in mean metallicities. 
We then consider two different ways of comparing the DG streams to surviving dwarfs, assuming that present-day satellite galaxies are good proxies for progenitors of DG streams \citep{Panithanpaisal2021}.  First, we compare the mean metallicities of the DG streams with known satellite galaxies of the MW \citep{Simon:2019}, and list galaxy analogs in Table \ref{table:stream_progenitor}. In particular, the progenitors of the Orphan-Chenab and Jhelum ($\feh\sim-1.8$) streams are expected to be similar to Leo II and Carina \resub{in luminosity} ($M_V\sim-9.5$), the progenitors of the Indus and Palca streams ($\feh\sim-2.0$) are expected to be similar to Draco and Sextans \resub{in luminosity} ($M_V\sim-8.9$), and the progenitors of the Elqui and Turranburra streams ($\feh\sim-2.2$) are expected to be similar to Ursa Major I  \resub{in luminosity} ($M_V\sim-5$).  Second, we estimate the progenitor luminosities using the luminosity-metallicity relation from \citet{Simon:2019}, updated from \citet{Kirby:2013}, shown on the upper axis of the right panel in Figure \ref{fig:feh}. Our DG streams' progenitors are at the faint end of the classical dSphs and the bright end of the UFDs.

Adopting the progenitor dynamical mass estimates based on stream widths from \citet{Shipp:2018}, the mass-to-light ratios of the progenitors of these DG streams are around 20-200 \msun/\lsun, further supporting the conclusion that their progenitors are dark matter-dominated DGs.

We compare the progenitor luminosities derived from metallicity with the present-day stellar mass of some streams presented in \citet{Shipp:2018}, assuming a $M_*/L_V = 1.6$ \citep{Kirby:2013}. We find that the present-day stellar mass is lower than the progenitor mass. For example, the present-day stellar mass for Elqui stream is about 50\% of the estimated progenitor stellar mass, while for Indus stream it is only $\sim$10\% of its estimated progenitor mass. 
This indicates that current photometric surveys are incomplete in their areal coverage of these streams (e.g., Chenab as part of the Orphan-Chenab stream) and/or in their achieved surface-brightness limit. Alternatively, it is possible that some stars in the stream are sufficiently phase mixed and thus difficult to detect above the background of the stellar halo \citep[e.g.][]{Johnston:1998,Helmi1999}.

\subsection{Streams in Phase Space}

With RVs from \SSSSS, PMs from \gaia EDR3, and distances derived in Section \ref{sec:distance} for individual member stars, we now are able to study these streams in phase space.  In this paper, we adopt the MW potential from the best-fit parameters of \citet{McMillan:2017}, which include six components: bulge, dark matter halo, thin and thick stellar disk, and HI and molecular gas disks. 
We compute the following potential-dependent quantities for each stream member using \code{gala} v1.3 \citep{gala, adrian_price_whelan_2020_4159870} and \code{galpot} \citep{Dehnen:1998}: the total orbital energy $E_\mathrm{tot}$, pericenter $r_\mathrm{peri}$, apocenter $r_\mathrm{apo}$, eccentricity, and orbital period. We also compute potential-independent quantities: angular momentum $L_Z$ and $L_{XY}$ and orbital poles $l_\mathrm{pole},b_\mathrm{pole}$. We take the Sun’s position and 3D velocity from \code{astropy} \citep{astropy, astropy:2018} v4.0 with $d_\sun=8.122$\,kpc \citep{Gravity2018} and $\emph{v}_\sun=(12.9, 245.6, 7.78)$\,\kms \citep{Drimmel2018, Gravity2018,Reid2004}. 

\begin{figure*}[!htb]
    \centering
    \includegraphics[width=0.9\textwidth]{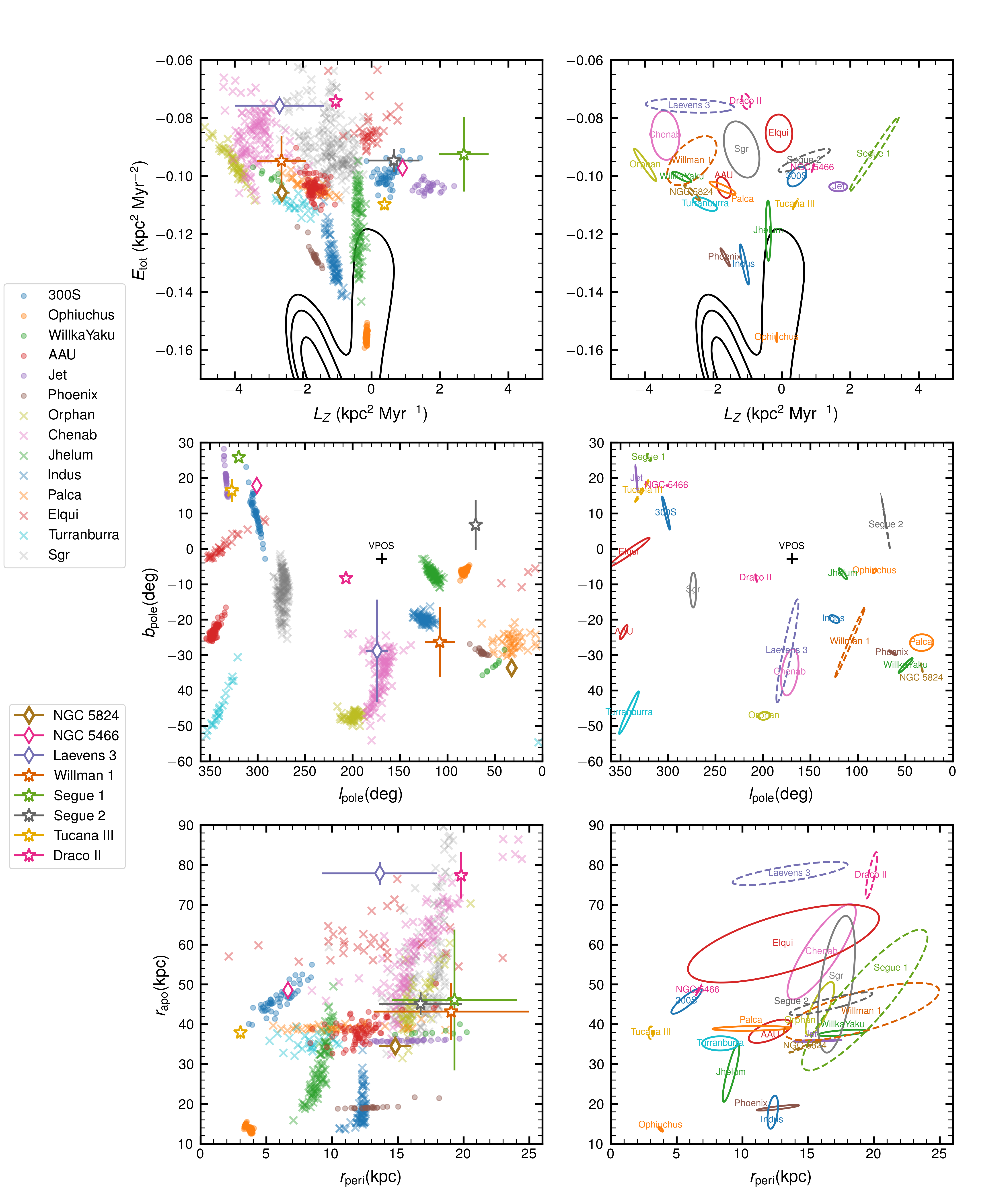}
    \caption{
    Orbital energy $E_\mathrm{tot}$ and angular momentum $L_Z$, where positive $L_Z$ corresponds to a retrograde orbit (top row), orbital poles $l_\mathrm{pole}$, $b_\mathrm{pole}$, where positive $b_\mathrm{pole}$ corresponds to a retrograde orbit (middle row), and orbital pericenters and apocenters (bottom row), computed from the spectroscopic member stars of each stream. The left column shows the scatter plot for the full sample of individual stream members, with circles for GC streams and crosses for DG streams.  To minimize crowding, only 10\% of the Sgr stream members are shown. Also shown are GCs (DGs) that might be associated with streams, with diamond (star) symbols for the median values and errorbars corresponding to one standard deviation. In the right column,  ellipses with solid (dashed) lines present 1-sigma confidence levels for the streams (DGs or GCs). Note that some ellipses are very small. In the top row, the black contours show the density distribution of all \SSSSS stars, which is dominated by the disk (negative $L_Z$) and GES ($L_Z\sim0$).
    In the middle row, the black plus sign shows the orbital pole of the ``Vast Plane of Satellites" (VPOS) from \citet{Pawlowski2013}.
    }
    \label{fig:orbit}
\end{figure*}

We present the phase space distribution of the dozen streams in Figure \ref{fig:orbit}. In the left panels, each symbol represents a single stream member star, with circles for GC stream members and crosses for DG stream members. In the right panels, we use an ellipse to show the 1-sigma contour assuming the distribution is a 2D-Gaussian. From top to bottom we show $E_\mathrm{tot}-L_Z$, orbital poles and peri/apocenters, respectively. Table~\ref{table:stream_orbital} summarizes the orbital properties, which are computed as medians across member stars in each stream.

Two GC streams, 300S and Jet, have a positive $L_Z$ and are therefore on retrograde orbits.\footnote{We adopt a right-handed coordinate system, in which a positive $L_Z$ corresponds to a retrograde orbit.} Three streams -- Ophiuchus, Jhelum, and Elqui -- \resub{have a negative $L_Z$, and a much smaller $|L_Z|$ compared to $L_{XY}$; hence we classify these three streams on near-polar orbits}. The remaining seven streams are on prograde orbits.
Except for Ophiuchus, all streams in our sample have Galactocentric distances larger than 10 kpc. All the streams have pericenters less than 20 kpc, which is well-matched with the fact that \resub{their progenitors were} tidally disrupted.

We note that the large scatter in Figure~\ref{fig:orbit} is mostly a result of measurement uncertainties, largely in the proper motion components. However, the large deviation between Orphan and Chenab is a result of the strong perturbation by the LMC \citep{Erkal2019}.

\begin{deluxetable*}{l r r r r r r r r r r r r}[!htb]
\tablecolumns{13}
\tablewidth{0pt}
\tabletypesize{\scriptsize}
\tablecaption{Orbital Properties of One Dozen Streams Observed with \SSSSS }\label{table:stream_orbital}
\tablehead{\\
Stream   & Orbit       &  $E_\mathrm{tot}$    &  $L_Z$  & $L_{XY}$& $l_\mathrm{pole}$ & $b_\mathrm{pole}$ &   $r_\mathrm{gal}^{a}$   &  $r_\mathrm{peri}$   &  $r_\mathrm{apo}$     &   Eccentricity  &Period   &   Possible \\
              &             &   (kpc$^2$/Myr$^{2}$) & (kpc$^2$/Myr) & (kpc$^2$/Myr) & (\degr) &(\degr) & (kpc)     &  (kpc)  &  (kpc)          &                  & (Myr)  &    Association  
}
\startdata
        300S  &  Retrograde  &     -0.10  &   0.44  &  2.65  &  302.0  &   10.1 &  22.2  &      5.8  &   45.8  &  0.77  & 560  &   NGC 5466  \\
   Ophiuchus  &       Polar  &     -0.16  &  -0.15  &  1.35  &   82.4  &   -6.2 &   4.3  &      3.8  &   13.7  &  0.57  & 180  &             \\
  WillkaYaku  &    Prograde  &     -0.10  &  -2.93  &  4.63  &   49.7  &  -32.9 &  36.1  &     17.6  &   37.9  &  0.37  & 570  &   NGC 5824  \\
         AAU  &    Prograde  &     -0.10  &  -1.70  &  3.90  &  346.5  &  -23.5 &  26.1  &     12.1  &   38.2  &  0.51  & 520  &      Palca  \\
         Jet  &  Retrograde  &     -0.10  &   1.65  &  4.68  &  333.1  &   19.9 &  33.9  &     15.7  &   35.9  &  0.39  & 530  &             \\
     Phoenix  &    Prograde  &     -0.13  &  -1.66  &  2.94  &   63.6  &  -29.3 &  18.9  &     12.7  &   19.1  &  0.20  & 320  &             \\
      Orphan  &    Prograde  &     -0.10  &  -4.00  &  3.67  &  198.7  &  -47.1 &  17.6  &     15.9  &   43.9  &  0.47  & 630  &            \\
      Chenab  &    Prograde  &     -0.09  &  -3.41  &  4.77  &  171.7  &  -34.8 &  28.0  &     16.0  &   58.2  &  0.56  & 800  &  Laevens 3  \\
      Jhelum  &       Polar  &     -0.12  &  -0.40  &  3.14  &  115.8  &   -7.0 &  10.8  &      9.1  &   27.8  &  0.51  & 380  &             \\
       Indus  &    Prograde  &     -0.13  &  -1.10  &  3.04  &  124.8  &  -19.8 &  12.4  &     12.3  &   17.9  &  0.19  & 300  &             \\
       Palca  &    Prograde  &     -0.10  &  -1.74  &  3.53  &   32.8  &  -26.5 &  38.5  &     10.7  &   39.0  &  0.57  & 520  &        Cetus, AAU  \\
       Elqui  &  Polar  &     -0.09  &  -0.09  &  5.18  &  339.0  &   -0.9 &  51.0  &     13.1  &   60.3  &  0.65  & 810  &             \\
 Turranburra  &    Prograde  &     -0.11  &  -2.25  &  2.26  &  340.8  &  -46.3 &  31.7  &      8.3  &   35.2  &  0.62  & 460  &             \\
\enddata
{\footnotesize \tablecomments{ 
$^{a}$ Distance to the Galactic center, median of all stream members.
}}
\end{deluxetable*}

\subsubsection{Association with MW Satellites}

In order to study the possible associations between the streams and other MW satellite systems, we also compute the orbital properties of the MW's GCs and DGs as we did for the streams, using 6D information from \citet{vasiliev_GCs} for GCs and Pace et al. (in prep) for DGs, treating each system as a single test particle. In order to compute the uncertainties of the orbits, for each DG or GC we sample 1000 realizations from the reported uncertainties.

In Figure \ref{fig:orbit}, we present the GCs that overlap with \SSSSS streams in orbital space. We first check their locations on the $E_\mathrm{tot}-L_Z$ plane, as we expect the energies and angular momenta of systems from the same progenitor or the same group infall to be similar in the absence of external perturbations. We then check the orbital poles and peri/apocenters to confirm the association. The orbital poles might be slightly off (by a few degrees) as they are expected to drift between multiple wraps \citep[e.g.][]{Erkal:2016a}. From this comparison, we find that three GCs (shown as open diamond symbols in the left panels of Figure \ref{fig:orbit}) have possible connections with the dozen streams: NGC 5824 and Willka Yaku, NGC 5466 and 300S, and Laevens 3 and Orphan-Chenab. We summarize these findings in Table \ref{table:stream_orbital}, with further discussion in Section \ref{sec:asso}.

For DGs, we limit our galaxy sample to those with median pericenter below 25 kpc and median apocenter below 90 kpc to better match the properties of the streams. These criteria result in a total of five UFDs: Segue 1, Segue 2, Willman 1, Draco II and Tucana III, shown as open star symbols in the same figure. Due to the large uncertainties in the orbital properties (mainly coming from the distance and PM uncertainties of the DGs), many of the DGs overlap with streams in phase space within the 1-sigma uncertainties. For example, Willman~1's orbital energy, angular momentum, pericenter and apocenter are all consistent with Willka Yaku, and Segue 1's orbital properties are similar to those of the Jet stream.
Among these UFDs, only Tucana III has clear tidal tails present \citep{Drlica-Wagner:2015, Li2019}, consistent with its extremely small pericenter ($\sim3$ kpc). The other four UFDs have no clear detection of tidal tails, although some evidence of tidal effects has been reported \citep[see, e.g.,][]{Kirby2013}. Given that these four DGs have pericenters similar to the other disrupted DG streams (i.e. $\lesssim 20$ kpc), further investigation might be needed to carefully examine these systems for possible tidal features. However, these DGs are all less luminous ($M_V > -3$), and therefore are usually smaller in half-light radius than the progenitors of the DG streams in our sample, which may explain why these galaxies are more resistant to tidal disruption even though they lie on similar orbits to the DG streams. It is also worth mentioning that these five UFDs have ambiguous classifications between DGs and GCs except for Segue 1, for which both a clear velocity dispersion and metallicity dispersion have been identified \citep[e.g.][]{Norris2010,Simon:2011}.

\section{Discussion} \label{sec:discuss}

\subsection{``Too Big to Fail" in Streams?} \label{sec:tbtf}

Cosmological simulations show that MW-mass dark matter halos have too many massive subhalos\footnote{More accurately, too many subhalos of high central density.} compared to the MW's actual satellites, which is known as the ``too big to fail" problem \citep{Boylan-Kolchin:2011, Boylan-Kolchin:2012}. We find a possibly similar too big to fail problem with the number of massive stellar streams in simulations.

Using the FIRE-2 suite of cosmological simulations, \citet{Panithanpaisal2021} studied the stellar streams formed from disrupted satellite galaxies in a suite of MW-mass galaxies. These simulations resolve stellar streams with progenitor stellar masses between $5\times 10^5 \msun$ and $10^9 \msun$, which are generally more massive than our DG stream progenitors. According to Figure \ref{fig:feh}, our most massive DG stream is Orphan-Chenab at $M_V\sim-9.1\pm1.4$, where the uncertainty comes from a scatter of 0.16 dex in \feh from the mass-metallicity relation \citep{Simon:2019}. This luminosity corresponds to a $M_* \sim 6^{+15}_{-4}\times 10^5 \msun$ assuming $M_*/L_V = 1.6$ \citep{Kirby:2013}, which is at the lower mass end of the streams in the FIRE-2 simulations. Except for the Sgr stream, none of the known MW DG streams have a stellar mass higher than the Orphan-Chenab stream\footnote{Although GES, the Helmi Stream, and other newly discovered stellar substructures \citep[e.g.,][]{Naidu2020} are likely from disrupted DGs, we do not consider them here as we only consider streams that are not phase-mixed, following the same definition as used in \citet{Panithanpaisal2021}}. However, cosmological simulations from 13 MW-like galaxies predict a total of 3-10 streams in this mass range, with a median of 8, which is much higher than the number inferred from observations.  If not considering the stellar mass, the other way to intepret the observational result is that we have not found any DG streams with a mean metalliciity between $\feh\sim-1.78$ (Orphan-Chenab) and $\feh\sim-0.5$ (Sgr). Two other DGs have shown possible tidal features, Antlia 2 and Crater 2; however, their metallicities are at $\feh = -1.77$ and $\feh=-2.10$, respectively \citep{Ji2021}. This is reminiscent of the ``too big to fail” problem among Milky Way satellites, and in fact the two may share a common origin.

A few possibilities might alleviate this tension. Firstly, it is possible that the simulations are overpredicting the number of massive streams, as the majority of these MW analogs ($M_{200} \sim 1.1-2.1 \times 10^{12}\msun$) are at the high end of the measured MW mass range, especially when the LMC is included in the fit. For example, \citet{Shipp2021} found that, when fitting the LMC mass with stellar streams, a lighter MW mass around $M \sim 0.8\times 10^{12}\msun$ fits the stream data better. The smaller number of observed massive DG streams relative to simulations could therefore be a hint of a smaller value of the MW mass.
Alternatively,  more massive DG streams in the MW halo may yet be undiscovered, either at larger distances that cannot be reached with current survey depths, or because the previous searches were biased toward metal-poor overdensity structures and missed the most metal-rich (and therefore massive) streams. Note that all DG streams presented in this paper were (partially) discovered in the DES data with a matched filter based on a metal-poor ($\feh\sim-1.9$), old (age = 13 Gyr) isochrone \citep{Shipp:2018}. 

\subsection{Prograde vs. Retrograde}\label{sec:prograde}

\citet{Panithanpaisal2021} showed that from FIRE-2 simulations we expect an even distribution of the satellite galaxies on prograde and retrograde orbits. However, as shown in Figure \ref{fig:orbit}, the majority of our streams are on prograde orbits with $L_Z < 0$ (or $b_\mathrm{pole}<0$). In particular, none of the DG streams are on retrograde orbits, except for Elqui, which is on a near-polar, slightly retrograde orbit. Although these DG streams' progenitors are all less massive than those shown in \citet{Panithanpaisal2021}, it is still striking that none of the DG streams from our sample are on a retrograde orbit. More intriguingly, as of now, none of the known DG streams in the MW are on retrograde orbits. 

\begin{figure}[!htb]
    \centering
    \includegraphics[width=0.45\textwidth]{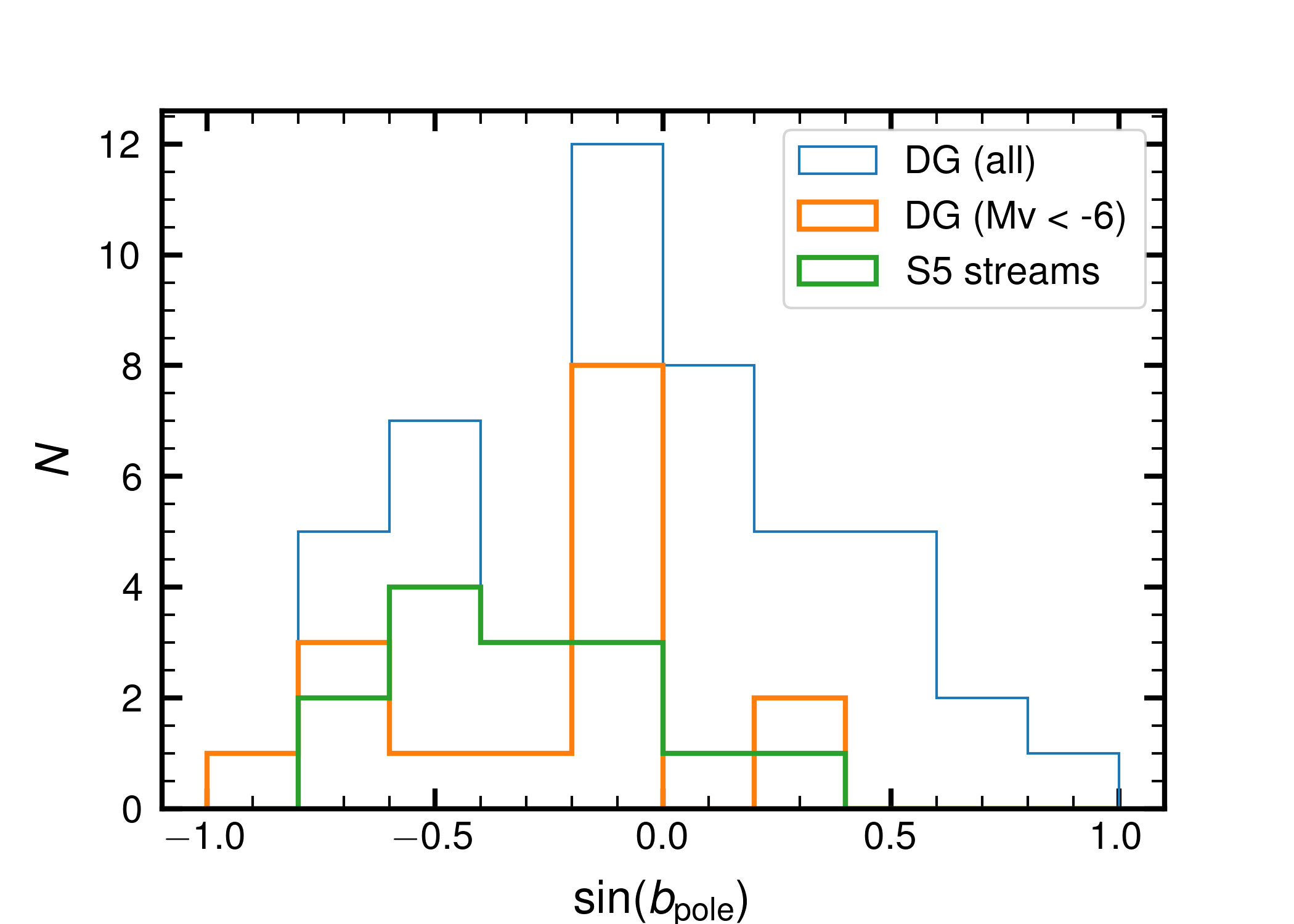}
    \caption{Distribution of $\sin(b_\mathrm{pole})$ for all MW DG satellites (blue), the brightest DG satellites (orange) and one dozen streams (green), where a positive (negative) $\sin(b_\mathrm{pole})$ corresponds to a retrograde (prograde) orbit.  The distribution for all DG satellites is reasonably symmetric. The large spike at $\sin(b_\mathrm{pole})\sim-0.1$ in the distribution of brightest DG satellites corresponds to the VPOS plane satellites. Both the massive DG satellites and the dozen streams have a preference for prograde orbits. }
    \label{fig:retrograde}
\end{figure}

In Figure \ref{fig:retrograde}, we compare the orbital orientation of the streams to relatively luminous DGs in the MW with $M_V < -6$, as this luminosity threshold is similar to the faintest DG streams in our sample, as derived from their metallicities. Intriguingly, among these 16 DGs, only two -- Fornax and Crater 2 -- are on retrograde orbits, showing a strong imbalance in prograde versus retrograde orbits, similar to the streams. However, many of these are DGs clustered in the ``Vast Plane of Satellites" \citep[VPOS;][]{Pawlowski:2012,Pawlowski:2015}, whose orbital pole is shown as the black `+' symbol in the top right panel of Figure \ref{fig:orbit}. Although both the streams and dSphs have a preference for prograde over retrograde orbits, the orbital poles of the streams do not seem to be clustered like the VPOS \citep{Riley:2020}.

The preference for prograde orbits in both streams and relatively luminous DGs may suggest that dwarf galaxies, especially the most massive DGs, were accreted onto the MW via group infall from one filament at a time, resulting in strong clustering in either prograde or retrograde orbits.
For example, the progenitors of some of these streams may have been accreted as a group early on with prograde orbits, while some of the relatively luminous DGs may have been accreted more recently from a single filament, resulting in a strong clustering in the VPOS plane.

\subsection{Eccentricities and Orbital Phase}\label{sec:phase}

\begin{figure*}[!htb]
    \centering
    \includegraphics[width=0.95\textwidth]{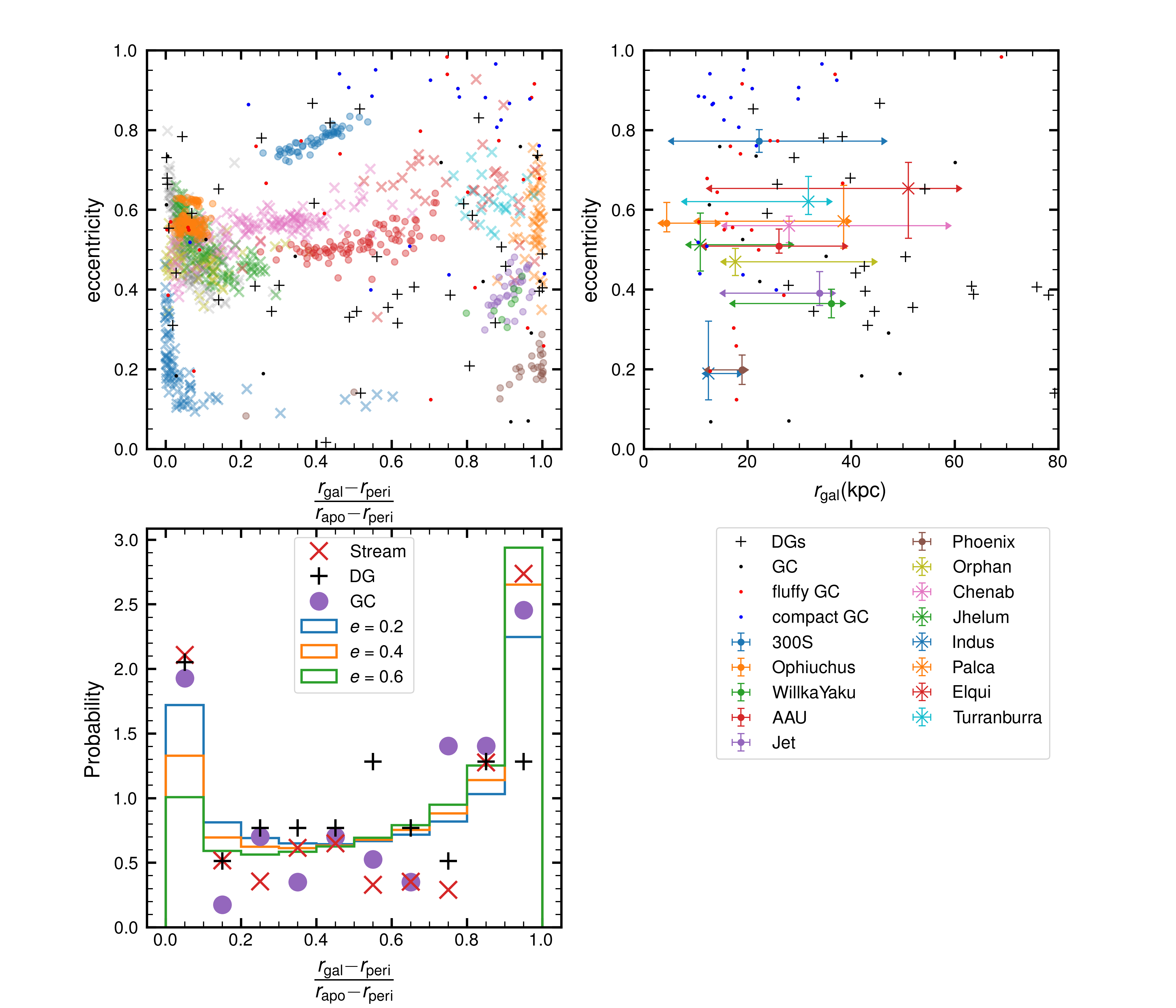}
    \caption{(upper left) Eccentricity $e = \frac{r_\mathrm{apo}-r_\mathrm{peri}}{r_\mathrm{apo}+r_\mathrm{peri}}$ vs ratio $f = \frac{r_\mathrm{gal}-r_\mathrm{peri}}{r_\mathrm{apo}-r_\mathrm{peri}}$. Each symbol corresponds to member stars in each stream. $f=0$ ($f=1$) indicates the star is close to the pericenter (apocenter). Also plotted are GCs (black dots) and DGs (black plus symbol). Only GCs with Galactocentric distances larger than 10 kpc are shown; only DGs with apocenters smaller than 300 kpc are shown. In addition, we highlight those ``fluffy" GCs in red and ``compact" GCs in blue, defined in \citet{Baumgardt2010} and \citet{Gieles2021}.  (upper right) Eccentricity vs Galactocentric distance ($r_\mathrm{gal}$) to the streams, where the circles and cross symbols indicate the current location of the stream and triangles shows the pericenter and apocenter of each stream. Each symbol corresponds to one stream, calculated as the median value of all stream members shown on the upper left. The errorbars on the eccentricities show the 16th and 84th percentiles from all stream members.(lower left) Histogram of ratio $f$ for stream, DG, and GC, respectively. Also plotted are the expected distributions for an outer halo star in an orbit with an eccentricity of 0.2, 0.4 and 0.6, respectively, under the same MW potential. The distributions predict a pile-up near $f\sim0$ and $f\sim1$, which matches what is seen in the GCs and streams. However, the current known DGs have a lack of pile-up at apocenter relative to predictions.}
    \label{fig:ecc}
\end{figure*}

To study the orbital phase and eccentricity ($e$) of the dozen streams, we define the ratio $f$ as a proxy to represent the orbital phase in the radial direction,  $f = \frac{r_\mathrm{gal}-r_\mathrm{peri}}{r_\mathrm{apo}-r_\mathrm{peri}}$, which was first used by \citet{Fritz:2018} to study the orbital phase of MW DGs; $f=0$ indicates that the stream is close to the pericenter and $f=1$ indicates that the stream is near apocenter.  Here $r_\mathrm{gal}$ is the current distance to the Galactic Center. In the upper panel of Figure \ref{fig:ecc}, we show the ratio $f$ and eccentricity for every stream member.
In the upper right panel, we show the median value for each stream, where the circle and cross symbols show the median distance to the Galactic Center and the arrows show the median pericenter and apocenter for each stream. Interestingly, among these dozen streams, four streams (Ophiuchus, Jhelum, Indus and Orphan-Chenab) are close to pericenter, while six streams (Palca, Turranburra, Elqui, Jet, Willka Yaku, and Phoenix) are close to apocenter; only AAU and 300S have $0.2<f<0.8$. 

We include GCs and DGs in Figure \ref{fig:ecc} for comparison. For GCs, we only include those with Galactocentric distance $r_\mathrm{gal} > 10$ kpc to match our streams, which we refer to as distant GCs in later discussion. In particular, we highlight those tidally filling (or ``fluffy")  GCs in red and compact GCs in blue \citep[see details on the definitions in][]{Baumgardt2010}, many of which also display tidal features \citep{Gieles2021}. For DGs, we consider DGs whose apocenters are less than 300 kpc, which is close to the virial radius of the MW.

We also show the probability distribution (normalized to one) of the ratio $f$ in the lower left panel of Figure \ref{fig:ecc}, for streams, DGs and GCs, respectively. For GCs and DGs, each DG and GC is counted as one occurrence. For streams, we count each stream as one occurrence (here Orphan and Chenab are considered one stream) but redistribute the weight to each stream member star evenly. Also plotted are the expected probabilities of $f$ for an outer halo star in an orbit with an eccentricity of 0.2, 0.4 and 0.6, respectively, under the same \citet{McMillan:2017} MW potential. We compute the distribution by sampling stars with similar orbits to the streams, DGs and GCs, but at particular eccentricities. The sampled distributions present pile-ups at both pericenter and apocenter in $f$, since $\frac{d}{dt}r_\mathrm{gal}$ is close to zero at near pericenter and apocenter. 
We see similar pile-ups in both pericenters and apocenters for the GCs and streams, indicating that there is no evidence for an observational selection effect reducing the observed number of streams or GCs near apocenter.
On the other hand, DGs show similar pile-ups around pericenter, but a deficiency near the apocenter of their orbits \citep[e.g.,][]{Simon2018, Fritz:2018}, indicating that a number of MW DG satellites near their apocenters may remain to be detected.

For the eccentricity, the upper panels show that the majority of streams are clustered around $e\sim 0.55$.
In particular, except for the Indus stream, all DG streams have an eccentricity between 0.45 and 0.65. The eccentricity distribution of streams is most similar to the ``fluffy" GCs, while compact GCs are more eccentric ($e\sim0.9$) and nearby ($r_\mathrm{gal}<40$ kpc), and DGs are less eccentric ($e\sim0.4$) and more distant ($r_\mathrm{gal}>40$ kpc), as shown in the upper right panel of Figure \ref{fig:ecc}. 

\subsection{Metallicity and Eccentricity of the GC streams}\label{sec:gc_feh}

\begin{figure}[!htb]
    \centering
    \includegraphics[width=0.45\textwidth]{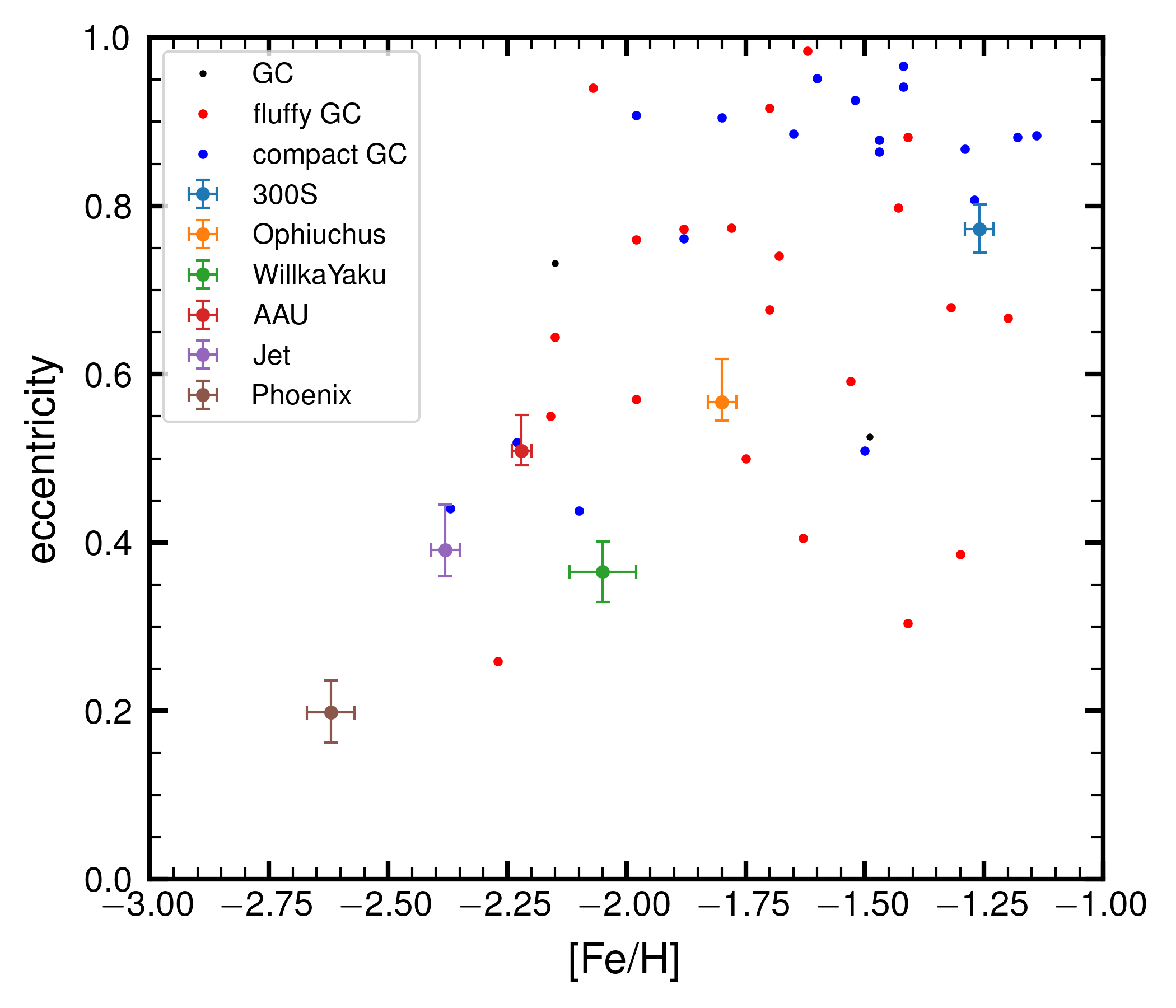}
    \caption{Eccentricity vs mean metallicity for all GC streams, showing a trend that more metal-rich streams are on higher eccentricity orbits. Also shown are known GCs whose Galactocentric distances are larger than 10 kpc. Metallicities of GCs are taken from \citet{Harris2010}. The GC streams have, in general, lower metallicities than GCs at similar distances. }
    \label{fig:feh_ecc}
\end{figure}

Figure \ref{fig:feh_ecc} shows that four of the six GC streams have a mean metallicity $\feh < -2$, while a majority of the intact distant GCs have a mean metallicity $> -2$, indicating that the progenitors of our GC streams are in general more metal-poor than the distant GCs. Such a difference in metallicity may point to a different origin for the GC streams compared to most of the distant GCs (e.g., accreted versus in-situ formation). Alternatively, if both populations are accreted, the disrupted GC streams may have been accreted earlier, and therefore their progenitors could have been more metal-poor. The latter possibility is also consistent with the fact that no clear progenitors are seen for any of the streams in our GC stream sample.

Even more interestingly, we find that these six GC streams show a correlation between the mean metallicity and eccentricity of the stream, where more metal-rich streams are on orbits with higher eccentricity. As these streams' progenitors are likely GCs accreted together with their parent galaxies, the \resub{GC metallicity may be correlated with the metallicity (and thus the mass) of the parent galaxy. The GC metallicity distribution function (MDF) may not match the parent galaxy MDF exactly, but it is clear that high-metallicity GCs must come from more massive parent galaxies.}
Such a correlation could therefore be explained in two scenarios. The first explanation is that the larger eccentricity GC streams are usually associated with a massive (and therefore more metal-rich) and recent merger, while the low eccentricity GC streams are accreted earlier \resub{with relatively metal-poor galaxies}.
Recent simulations also show a trend between the accretion time and eccentricity, where high eccentricity stars are likely associated with a recent merger event \citep{Mackereth2018}. Alternatively, assuming all these GCs' parent galaxies were accreted at a similar time, then as a result of dynamical friction, the orbits of the more massive galaxies \resub{tend to} become more radial and eccentric \resub{faster than lower-mass galaxies (although the details depend on other factors like the initial orbital eccentricity; \citealt{Vasiliev:2021b})}. 

We also show the eccentricity and metallicity of distant GCs in the same panel. Most of these distant GCs have a higher metallicity and also a higher eccentricity, and show a weaker metallicity-eccentricity correlation than our GC streams. 

We note that such a relation in the streams could just be a coincidence due to the small sample size. The relation is largely driven by the two most extreme streams.  300S, the most metal-rich stream in our sample, is possibly associated with a very radial merger, GES, which was likely accreted $\sim10$ Gyr ago \citep[e.g.][]{Belokurov:2018,Helmi:2018,Naidu2021}.
On the other hand, the Phoenix stream has a metallicity lower than any known GC \citep{Wan2020}, and also seems to have a more circular orbit than any known distant GC. As the dissolution time of a GC stream is anti-correlated with the eccentricity \citep{Baumgardt2003}, Phoenix’s progenitor would have a maximum lifetime possible in its orbit. This may explain why it is not fully mixed yet, even though it might have fallen into the MW a long time ago given its extremely low metallicity.

Finally, we note that we do not see such correlation in our DG streams since the metallicity range for DG streams is very narrow, as shown in the right panel of Figure \ref{fig:feh}.

\subsection{Stream Velocity Dispersions}\label{sec:largedispersion}

\begin{figure*}[!htb]
    \centering
    \includegraphics[width=0.45\textwidth]{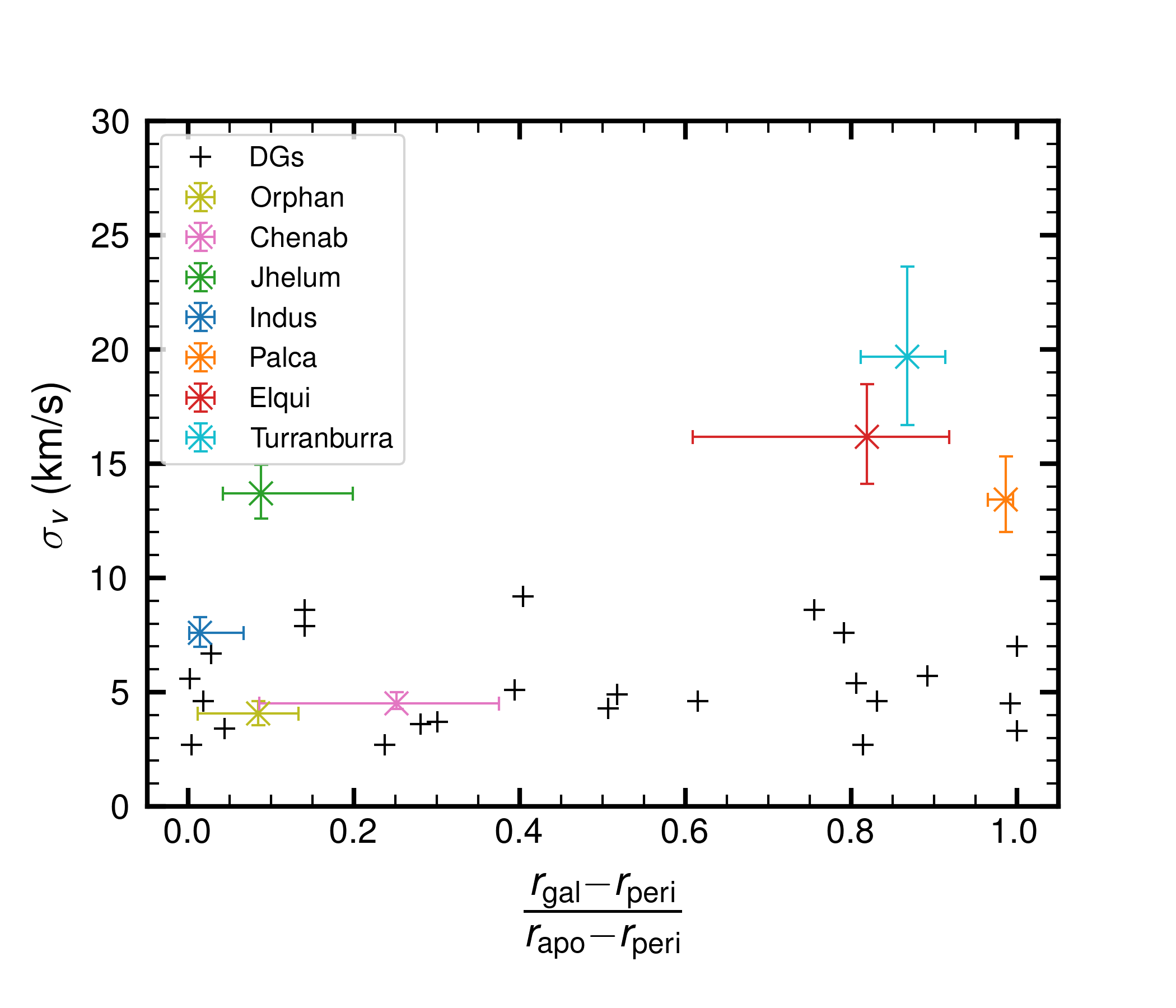}
    \includegraphics[width=0.45\textwidth]{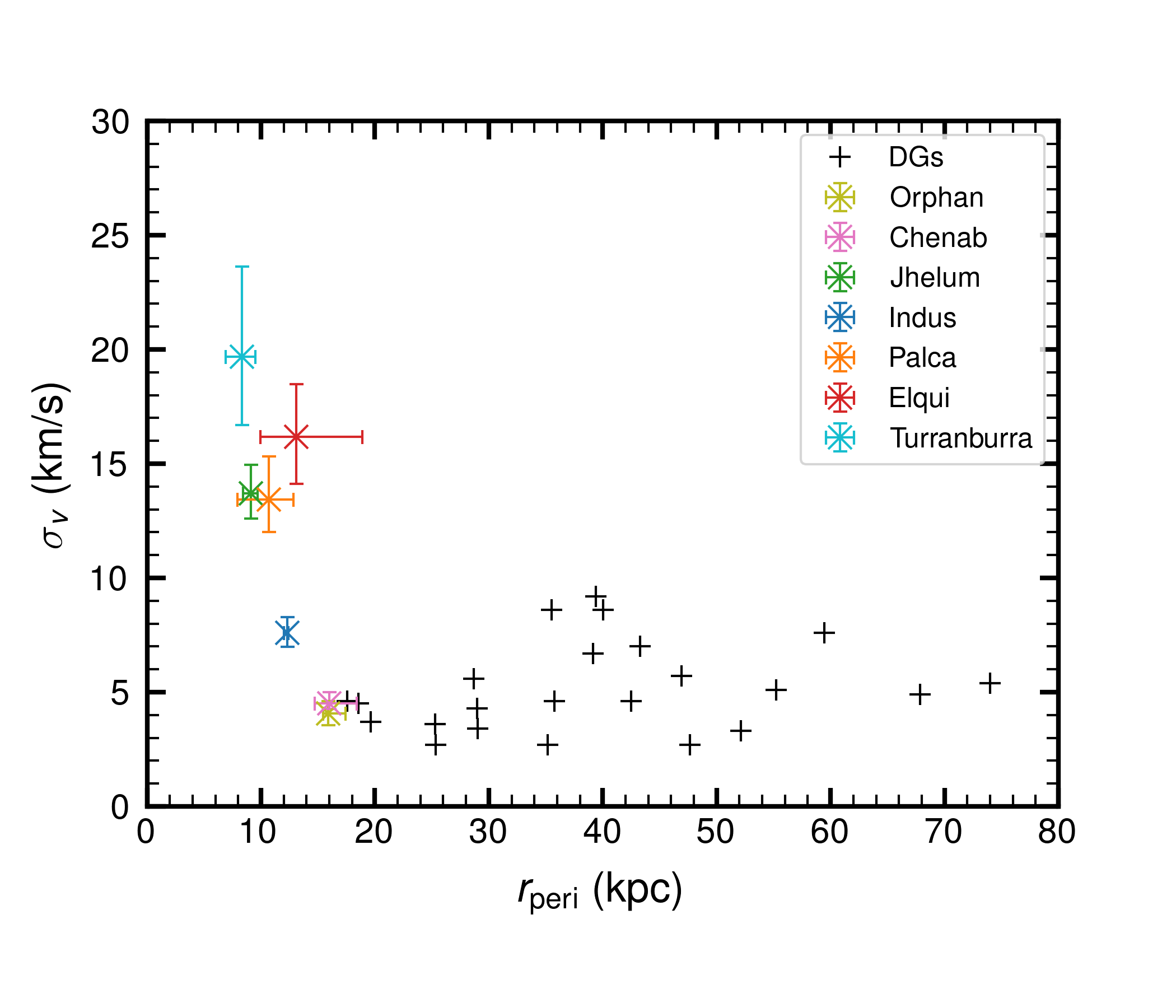}
    \caption{(left) Velocity dispersion of DG streams versus ratio $f =\frac{r_\mathrm{gal}-r_\mathrm{peri}}{r_\mathrm{apo}-r_\mathrm{peri}} $. The errorbars are computed from 16th and 84th percentile of all stream members.  Except for Jhelum, streams with large velocities dispersions are located near the apocenters of their orbits. (right) Velocity dispersion of DG streams versus their pericenters. Streams with the smallest pericenters tend to have larger velocity dispersions. For both panels, MW DG satellites with a resolved velocity dispersion are also shown, which mostly have velocity dispersions ${<}$10 \kms. We note that in the right panel, there appears to be a deficiency of DGs with $r_\mathrm{peri} < 20$ kpc. Actually, a few DGs or DG candidates -- such as Tucana III, Segue 2, Draco II, and Triangulum II -- have pericenters less than 20 kpc. However, as these systems only have upper limits for their velocity dispersions, they are not included in this plot.}
    \label{fig:sigmav}
\end{figure*}

DG streams are expected to have larger velocity dispersions than GC streams (Figure \ref{fig:feh}) due to the large dynamical mass of their progenitors' dark matter halos. However, Elqui, Jhelum, Palca and Turranburra show dispersions that are larger than 13\,\kms, which is higher than every MW satellite galaxy \citep{Simon:2019} except for the LMC and Small Magellanic Cloud (SMC). 
\resub{Since the inferred stellar masses of the progenitors of DG streams are smaller than most of the classical dSph satellites, and consevation of phase-space density would suggest that velocity dispersion of a stream should decrease with time \citep{Helmi1999}, the extremely large velocity dispersion suggests that additional processes have heated the streams.}
Interestingly, except for Jhelum -- whose large velocity dispersion may be associated with its complicated morphology \citep{Bonaca2019,Shipp2019} -- the other three streams with large velocity dispersions are all near apocenter, as shown in the left panel of Figure~\ref{fig:sigmav}. This suggests the dispersion of DG streams might be correlated with the stream's orbital phase. 
However, \citet{Panithanpaisal2021} found in the FIRE-2 simulations that streams near pericenter tend to have higher velocity dispersions, contradictory to what is shown in Figure~\ref{fig:sigmav}.

We also explore the relation between velocity dispersion and pericenter, shown as the right panel of Figure \ref{fig:sigmav}. The four DG streams with large velocity dispersions have relatively small pericenters compared to the Orphan-Chenab stream, which has the lowest velocity dispersion among all DG streams. In contrast, although the Indus stream has a small pericenter, its eccentricity is also much smaller than the other five DG streams. The large velocity dispersions observed for some DG streams may be explained by a combination of the eccentricity and pericenter of each stream. For globular cluster streams, such a relation would be expected since the streams with small pericenters and large eccentricities will have correspondingly smaller tidal radii \citep{kuepper2012} and thus larger velocity dispersions for the stripped stars. However, for the observed dwarf galaxies, the velocity dispersion does not strongly depend on the radius \citep[e.g.][]{Walker2007}. Thus, more theoretical work is needed to determine if this could explain the entirety of the effect.

Alternatively, \citet{Errani2015} argued that the internal dynamics of tidal streams are largely affected by the dark matter profile in which their progenitors are embedded; a larger velocity dispersion in the DG streams may indicate that the progenitors were embedded in a cored dark matter halo instead of a cuspy one. Our DG stream sample would be a valuable sample for such cusp/core dark matter profile studies.

On a similar note but different from the DG streams, \citet{Malhan2021b} proposed that the dynamical properties of accreted GC streams are also sensitive to the central dark matter density profile of their parent DGs --
GC streams are dynamically hotter if their parent DGs reside in a cuspy subhalo than if they reside in a cored subhalo. In their simulations, GC streams originating from a cuspy parent subhalo have velocity dispersions larger than 4 \kms. All our GC streams have velocity dispersions below 4 \kms, suggesting that if the modeling from \citet{Malhan2021} is correct and our GC streams are indeed accreted, the progenitors of these GC streams resided in cored subhalos.

\resub{Finally, we note that our inferred velocity dispersions could be inflated by non-members or binary star motions.
This is especially relevant for the dwarf galaxy streams, as the intrinsic metallicity spread makes it harder to remove foreground stars.
}

\subsection{Stream Associations with Other Objects}\label{sec:asso}

\resub{Dwarf galaxies can accrete with a population of their own globular clusters or dwarf galaxy satellites. We thus might expect to see kinematic associations between our streams and/or other halo objects. Conversely, it should be possible to associate GC streams with their parent dwarf galaxies, whether those galaxies are intact or tidally disrupted. We now discuss several likely associations.}

\subsubsection{300S, NGC5466, Tucana III and Gaia-Enceladus-Sausage (GES)}

\citet{Fu:2018} argued that the 300S progenitor was likely to be a compact DG, based on the chemical abundances of 300S member stars from APOGEE and SEGUE.
We do not detect a metallicity dispersion with a much larger sample of 300S member stars, and 300S has a small stream width of $\sigma_w = 0.4$\degr\ or $\sim 110$ pc (at a heliocentric distance of 16 kpc), similar to other GC streams.\footnote{$\sigma_w = 0.4$\degr\ corresponds to a FWHM stream width of 0.94\degr\ reported in \citet{Fu:2018}}
Thus, we argue that 300S's progenitor is more likely to be a GC than a DG. Furthermore, given its high metallicity ($\feh\sim-1.3$), if it is a DG, then it would have been very massive, according to the galaxy mass-metallicity relation. This is unlikely unless the progenitor is the compact core or the nucleated star cluster of a massive galaxy. Given its location on the $E_\mathrm{tot}-L_Z$ plane in Figure \ref{fig:orbit}, and its high eccentricity ($e=0.80$) and metallicity, we postulate that 300S's progenitor was highly likely to have been a GC of a past merger on a polar-to-retrograde orbit, very likely to be  GES \citep{Belokurov:2018, Helmi:2018}. Alternatively, 300S's progenitor may be the nucleated star cluster of GES; this picture might also match with the abundance pattern seen in \citet{Fu2019} from APOGEE. More high-resolution spectroscopic observations on 300S might solve this puzzle.

NGC 5466 overlaps with 300S in the $E_\mathrm{tot}-L_Z$ plane, and has almost identical orbital poles and peri-apocenters to 300S.
However, the metallicity of NGC~5466 \citep[$\feh = -1.97\pm0.13$;][]{Lamb2015} is significantly more metal-poor than 300S ($\feh\sim-1.3$), making it unlikely to be the direct progenitor of 300S. 
Similarly, Tucana III is also very close to 300S in the $E_\mathrm{tot}-L_Z$ plane with a high eccentricity, but its mean metallicity \citep[$\feh\sim-2.5$;][]{Simon:2017,Li2019} is also much lower. 
Both NGC 5466 and Tucana III show tidal tail features \citep{Drlica-Wagner:2015, Belokurov2006b}, although the progenitor cores are not yet fully disrupted.
It is possible that NGC 5466, Tucana III, and the progenitor of 300S are satellite GCs or UFDs from the same infall of their potential parent galaxy, GES. 

\subsubsection{Palca, Cetus, and AAU}

Although the Palca stream was discovered in DES \citep{Shipp:2018}, no specific observations were planned by \SSSSS \citep{Li2019} due to its extremely diffuse structure. However, \citet{Li2021} detected a second structure with a heliocentric velocity of $\sim100$\,\kms in the AAU field. Based on its location, PM, and heliocentric distance, \citet{Li2021} argued this kinematically cold structure is associated with the Palca stream. \citet{Chang2020} argued that Palca is likely the southern extension of the Cetus stream. Our Palca stream data match the kinematic prediction of \citet{Chang2020}, confirming that Cetus and Palca are indeed one stream. A more detailed study of this connection is discussed in \citet{Yuan2021}.

Although Palca member stars are identified in the AAU field, the PMs of the Palca members are almost perpendicular to those of the AAU stars. Their RVs also differ by $>$100\,\kms. Therefore, no connections were originally considered between these two streams. Instead,
\citet{Li2021} pointed out that AAU's orbit is very close to Whiting 1, NGC 5824, and Pal 12 in action space, which are themselves possibly associated with the Sgr dwarf according to \citet{Massari:2019}.  \citet{Li2021} therefore suggested that the progenitor of the AAU stream might have been accreted with the Sgr dwarf.

Interestingly, we find here that the AAU stars and Palca stars exactly overlap in the $E_\mathrm{tot}-L_Z$ plane (upper panels of Figure \ref{fig:orbit}) and show very similar peri/apocenters (lower panels of Figure \ref{fig:orbit}). The orbital poles of the two streams are also very close (middle panels), but are slightly different in longitude ($l_\mathrm{pole}$), which is expected between multiple wraps in a non-spherical potential. We therefore argue that AAU is more likely to be associated with the Palca/Cetus stream than with the Sgr stream, and is likely on a different wrap from Palca, which leads to a small drift in orbital pole.
Since the metallicity mean and dispersion of the two streams are slightly different, it is unlikely that the two streams share the same progenitor. However, it may be that AAU's progenitor was a GC of Palca's progenitor. This possibility would be exciting because no GCs have been found around DGs at similar metallicities to Palca ([Fe/H] = $-2.0$; e.g., Draco, Sextans, etc.).

We also note that several studies have suggested that NGC 5824 might be associated with the Cetus (i.e., Palca) stream \citep[e.g.,][]{Newberg:2009, Yuan2019}. \citet{Bonaca2021} also suggested that Willka Yaku might be related to the Cetus stream. Indeed Figure \ref{fig:orbit} shows that Palca/Cetus, AAU, NGC 5824, and Willka Yaku are all very close in phase space, and are likely all from the same group infall.
However, based on the $E_\mathrm{tot}-L_Z$ and poles, we conclude that Palca and AAU are more closely connected, and Willka Yaku and NGC 5824 are more closely connected (see next Section). 

\subsubsection{NGC 5824, Willka Yaku, Turbio and Triangulum/Pisces}

\begin{figure}[!htb]
    \centering
    \includegraphics[width=0.45\textwidth]{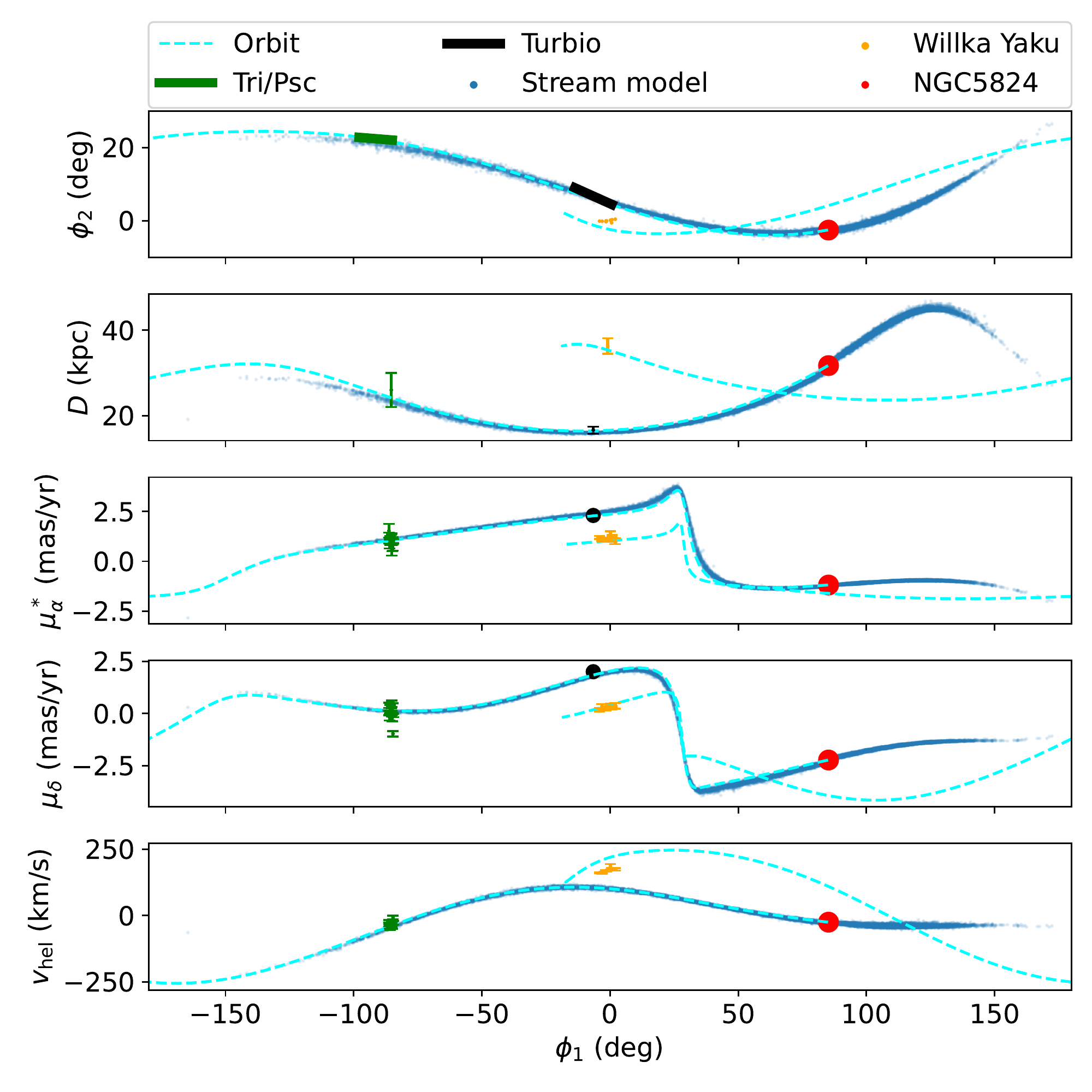}
    \caption{Comparison between the stream model of the NGC 5824, the orbit of NGC 5824 and the Turbio, Triangulum/Pisces and Willka Yaku streams. The blue particles show the predicted stream of NGC 5824 and the dashed cyan line shows the future orbit of NGC 5824. The red circle shows the location of NGC 5824. The green, black, and yellow symbols show the observables of Tri/Psc, Turbio, and Willka Yaku respectively. The coordinate system $\phi_1$ and $\phi_2$ are defined in Willka Yaku stream coordinates, using the transformations defined in \citet{Shipp2019}.}
    \label{fig:ngc5824}
\end{figure}

As shown in Figure \ref{fig:orbit}, Willka Yaku and NGC 5824 have very similar energies, angular momenta and orbital poles, suggesting they may have a common origin. 
\citet{Kuzma2018} found that NGC 5824 is remarkably extended in size, which might be an indication of tidal stripping.
\citet{Bonaca2021} argued that NGC 5824 may be the progenitor of both the Triangulum and Turbio streams. 
We therefore study the connections between these three streams and NGC 5824, shown in Figure \ref{fig:ngc5824}.

The Triangulum stream was first discovered by \citet{Bonaca:2012} in photometric data from the Sloan Digital Sky Survey Data Release 8 (SDSS DR8) at a distance of $26\pm4$ kpc. Shortly thereafter, \citet{Martin2013} found a kinematically cold structure in a similar part of the sky with spectroscopic data from SDSS DR8, which they dubbed the Pisces Stellar Stream. Here we refer to this structure as the Triangulum/Pisces (or Tri/Psc) stream. In Figure \ref{fig:ngc5824}, we show the on-sky position and distance from \citet{Bonaca:2012}\footnote{Note that \citet{Martin2013} found the distance of Pisces stream to be $35\pm3$ kpc. We did not use this distance because the distance from \citet{Bonaca:2012} is a better fit to our model.}, and PMs and RVs of likely members found by \citet{Martin2013}, cross-matched with \gaia EDR3 and SDSS. Observations of Turbio are planned but have not yet been conducted by \SSSSS; we therefore show the on-sky position and distance from \citet{Shipp:2018} and proper motion from \citet{Shipp2019}.

In order to explore the predicted debris of the NGC5824 globular cluster, we use the modified Lagrange Cloud stripping method \citep{Gibbons2014} as implemented in \cite{Erkal2019} to include the effect of the LMC. For the Milky Way potential, we use a realization drawn from the posterior chains in \cite{McMillan:2017}, which \cite{Shipp2021} gave a good match to 7 Milky Way streams (see Table A3 of \citealt{Shipp2021} for the potential parameters). We model the LMC as a Hernquist profile \citep{Hernquist1990} with a mass of $1.5\times10^{11}$ and a scale radius of 17.13 kpc, which matches the rotation curve measurements of the LMC at 8.9 kpc \citep{vandermarel2014}. For the present-day position and velocity of the LMC we use measurements of its proper motion \citep{kallivayalil2013}, distance \citep{pietrzynski2013}, and radial velocity \citep{vandermarel2002}. We model the progenitor of NGC 5824 as a Plummer sphere matching the observed mass and half-light radius \citep{baumgardt2018}.

The blue dots in Figure \ref{fig:ngc5824} show the stream model of NGC 5824 (position and velocity taken from \citealt{vasiliev_GCs}) stripped for 6 Gyr. The stream particles match with Turbio and Tri/Psc perfectly, but do not go through Willka Yaku. We note that if the LMC is ignored, the agreement of the NGC 5824 stream with Tri/Psc and Turbio is worse. Next, we consider whether Willka Yaku is consistent with a past or future orbit of NGC 5824 including the LMC. This is done by rewinding NGC 5824 for 6 Gyr in the presence of the Milky Way and LMC. At this time, the LMC is far from the Milky Way ($\sim 425$ kpc in the models used here) and the inner regions of the Milky Way can approximately be thought of as isolated. Particles are then placed along this past orbit (between 4-6 Gyr ago with a spacing of 0.5 Myr), initialized at the same time of 5 Gyr in the past, and integrated to the present. If there were no LMC, these particles would lie along the orbit of NGC 5824, but with the LMC, these particles will instead represent the original orbit of NGC 5824 before the LMC's infall. The blue dashed line shows the leading orbit of NGC 5824 using this technique. Member stars in Willka Yaku are closer to the orbit of NGC 5824 in a second wrap, but still have a small offset in the on-sky position and RV. We therefore conclude that, while NGC 5824 is likely the progenitor of Turbio and Tri/Psc, it is less likely the progenitor of Willka Yaku. However, given how similar their orbits are, it is very likely that the progenitors of Willka Yaku and NGC 5824 fell into the MW together.

Although Willka Yaku sits close to the future orbit of NGC 5824, the stream model does not reach its location despite being disrupted for 6 Gyr. Interestingly, Willka Yaku appears to sit at a lower energy than NGC~5824 (see Fig. \ref{fig:orbit}), which is inconsistent with it being in the leading debris of NGC 5824. However, there are significant uncertainties in this energy, and the energy may also have changed significantly due to perturbations since they were accreted, e.g., by Sgr, the LMC, or other galaxies.  

The metallicities for these systems are also very similar. NGC 5824's metallicity is $\feh =-1.94\pm0.10$ \citep{Roederer2016}. We find Willka Yaku's metallicity to be $\feh = -2.05 \pm 0.07$. \citet{Martin2013} reported a spectroscopic metallicity of $\feh = -2.2$ for Tri/Psc. 

\citet{Roederer2016} found that NGC 5824 showed internal heavy element abundance variations. It will therefore be interesting to search for similar abundance patterns in Willka Yaku, Turbio, and Tri/Psc with high-resolution spectroscopy.

\subsubsection{Indus, Jhelum and LMS-1}

\citet{Bonaca2019} suggested that Jhelum and Indus are tidal debris from the same DG at different orbital phases based on their 3D orbits. Our calculations show that the orbital properties of these two streams are close but quite distinct. Furthermore, their mean metallicities differ by more than 1.5$\sigma$. We therefore conclude that the two streams are unlikely to share the same progenitor, unless the progenitor possessed a significant metallicity gradient, causing a metallicity variation along the stream.

\citet{Malhan2021} argued that the LMS-1 stream \citep{Yuan2020} may be associated with the Indus stream based on their orbital properties. Our metallicity measurements for Indus ($\feh\sim-2.0$) are similar to those for the LMS-1 stream ($\feh\sim-2.1$), but Indus shows a significantly smaller velocity dispersion ($\sim7.5$\,\kms for Indus and $\sim20$\,\kms for LMS-1 from \citealt{Malhan2021}). This is consistent with the suggestion of \cite{Malhan2021} that the progenitor of Indus was a satellite galaxy of the progenitor of LMS-1.

\subsubsection{Orphan-Chenab and Laevens 3}
The Laevens 3 GC \citep{Laevens2015} overlaps with the Chenab stream in phase space. \citet{Longeard2019} found a mean metallicity of $\feh = -1.8 \pm 0.1$ for Laevens 3, which is consistent with Orphan-Chenab's metallicity. 
We therefore tentatively suggest that Laevens 3 is likely a GC that was accreted along with the Orphan-Chenab stream's DG progenitor. However, due to the potential influence of the LMC, more sophisticated modeling is necessary to confirm this association. We leave this detailed study to Koposov et al. (in prep).

\subsubsection{Clusters in the smallest galaxies}
Among the MW satellites, the LMC, SMC, Fornax, and the Sgr dSph are known to have their own GCs. However, lower mass DGs such as Sculptor, Carina, Sextans, and Draco have no GCs. With the exception of the Eridanus II DG at $\sim 380$\,kpc  \citep{bechtol15, koposov15}, none of the galaxies with luminosities below Fornax ($M_V \sim -13.5$) are known to host GCs. However, in this work, we found several possible associations that suggest that some Draco-to-Carina-like galaxies may have possessed GCs, including AAU's progenitor in Palca, and Laevens 3 in Orphan-Chenab. If any of these associations are real, they would move the boundary for the formation of GCs in the DGs to lower luminosities \citep[e.g.,][]{Kruijssen2020Natur,Wan2020} and may also affect our understanding of the specific frequency of GCs in DGs \citep[e.g.,][]{Huang2021}.

\section{Conclusions}\label{sec:conclusion}

We report the orbital and chemical properties of twelve stellar streams that have a robust detection of spectroscopic members in the current \SSSSS data set. Three other streams were observed by \SSSSS with no clear spectroscopic member detection. None of these streams have a clear progenitor embedded in the stream. Using line-of-sight velocities from \SSSSS, PMs from \gaia EDR3, and distances derived from BHB and RRL tracers, we summarize the properties of the streams in Table \ref{table:stream_progenitor} and Table \ref{table:stream_orbital}, and draw the following conclusions for these streams:

\begin{itemize}
\item The velocity dispersions and metallicity dispersions show that half of these streams have DG progenitors, while the other half originate from disrupted GCs (Figure \ref{fig:feh}). 

\item Our stream sample shows a significantly higher percentage of streams on prograde orbits than on retrograde orbits (Figure \ref{fig:orbit} and Figure \ref{fig:retrograde}). Out of the dozen streams, seven are on prograde orbits, three are on polar orbits, and the remaining two streams are on retrograde orbits. The most luminous DGs in the MW show a similar bias towards prograde orbits; only two out of the 16 most luminous MW satellites are on retrograde orbits. The fact that both massive DGs and our stream sample show a preference for prograde orbits may suggest that groups of massive galaxies have been accreted onto the MW through group infall, resulting in a non-uniform distribution of the orbital poles (Section \ref{sec:prograde}). 

\item For streams from disrupted DGs, the mean metallicities range from $\feh=-2.2$ to $-1.8$. The corresponding luminosities of the progenitors range from $M_V\sim-6$ to $M_V\sim-10$, using the mass-metallicity relation from \citet{Kirby:2013}. The fact that none of these DG streams are in a similar mass range to those in the FIRE-2 simulation, may indicate that the ``the too big to fail" problem observed in MW satellite galaxies extends to stellar streams (Section \ref{sec:tbtf}).

\item For streams from disrupted GCs, the mean metallicities have a much wider range, from $\feh=-2.7$ to $-1.2$. Our GC streams are in general more metal-poor than MW GCs at similar distances (Figure \ref{fig:feh_ecc}), suggesting that either the progenitors of the GC streams have a different origin or a different accretion history from the distant MW GCs that are not yet fully disrupted. We also find a clear trend from six GC streams that more metal-rich streams lie on more eccentric orbits (Section \ref{sec:gc_feh}). 

\item We compute a ratio $f = \frac{r_\mathrm{gal}-r_\mathrm{peri}}{r_\mathrm{apo}-r_\mathrm{peri}}$ as an analog of the orbital phase of the streams in radial direction, and find that 50\% of the streams are near apocenter and 30\% of the streams are near pericenter, matching expectations of apocenter and pericenter pileup (Figure \ref{fig:ecc}). We compare with GCs and DGs in the MW and find that although the pile-up at pericenter for DGs matches with the streams and GCs as well as expectation, there is a clear lack of DGs at apocenter. In addition, the eccentricities of the streams are mostly similar to those of the fluffy GCs in the MW, higher than the DGs and lower than the compact GCs at similar distances (Figure \ref{fig:ecc} and Section \ref{sec:phase}).

\item Four DG streams show large velocity dispersions ($\gtrsim 10$ \kms), which may result from a combination of high eccentricity and small pericenter of the stream orbits (Section \ref{sec:largedispersion}). On the other hand, all our GC streams show a velocity dispersion below 4 \kms.  

\item We compare the orbital properties of the streams to those of DGs and GCs in the MW, finding several possible associations (Section \ref{sec:asso}). In particular, the AAU stream's progenitor might be a GC of the Palca stream; NGC 5824 and the progenitor of the Willka Yaku stream may have fallen into the MW together; Laevens 3 might be a GC of the Orphan-Chenab stream; and NGC 5466, Tucana III and the progenitor of 300S may have fallen into the MW together with a massive merger, likely GES. We emphasize that these associations are based on the locations of these substructures in the $E_\mathrm{tot}-L_Z$ plane (Figure \ref{fig:orbit}). Further modeling work and/or observations are necessary to confirm these connections. Finding GCs or GC streams that are associated with DG streams will inform our understanding of the formation of GCs in the smallest galaxies.

\end{itemize}

This paper mainly focuses on the properties of the dozen streams observed in \SSSSS as our first results, serving as a starting point to understand the distant ($\gtrsim 10$ kpc) stream population in the MW halo. Recently, many other streams \citep[e.g.,][]{Ibata2019, Ibata2021} and substructures \citep[e.g.,][]{Naidu2020} have been discovered with \gaia and other spectroscopic efforts. We defer a more thorough discussion of these connections to a future paper. 

In addition to the study of stream properties presented in this work, this stream sample is also a unique dataset that may be used to study small-scale perturbations. In particular, 300S and Jet are both GC streams on retrograde orbits, and are therefore ideal targets for stream density variation modeling in order to search for evidence of dark matter subhalo flybys. Furthermore, these dozen streams would also be an ideal sample to constrain the MW potential to high precision \citep{Bonaca2018}. However, we note that many of these streams are strongly affected by the LMC which will be essential to model in such fits \citep[e.g,][]{Erkal2019,Shipp2021}.

\SSSSS has announced its first public data release \citep[DR1,][]{S5DR1}\footnote{DR1 is based on iDR1.5 and is now available at \url{https://zenodo.org/record/4695135}}, which contains all targets observed in 2018-2019 following the target selection, data reduction and survey validation described in \citet{Li2019}. The following streams mentioned in this paper are included in DR1: AAU, Elqui, Indus, Jhelum, Orphan, Chenab, Palca, Phoenix, Willka Yaku, Sgr, and Ravi (no detection). Observations taken in 2020-2021 are expected to be released in \SSSSS DR2 in late 2022.

\acknowledgments

\resub{The authors would like to thank the referee for detailed comments that helps clarification of the paper.} TSL would like to thank Ray Carlberg, Rohan Naidu, Ted Mackereth, Zhen Yuan, Nondh Panithanpaisal and Robyn Sanderson for helpful discussion. TSL also would like to thank Jiang Chang for sharing his Cetus model for comparison with our Palca data. TSL is supported by NASA through Hubble Fellowship grant HST-HF2-51439.001 awarded by the Space Telescope Science Institute, which is operated by the Association of Universities for Research in Astronomy, Inc., for NASA, under contract NAS5-26555.
ABP is supported by NSF grant AST-1813881.  
SLM, JDS, DBZ acknowledges the support of the Australian Research Council through Discovery Project grant DP180101791, and this research has also been supported in part by the Australian Research Council Centre of Excellence for All Sky Astrophysics in 3 Dimensions (ASTRO 3D), through project number CE170100013.
SLM and JDS are supported by the UNSW Scientia Fellowship program.
EB acknowledges support from a Vici grant from the Netherlands Organization for
Scientific Research (NWO).
APJ acknowledges a Carnegie Fellowship and support from the Thacher Research Award in Astronomy.

This paper includes data obtained with the Anglo-Australian Telescope in Australia. We acknowledge the traditional owners of the land on which the AAT stands, the Gamilaraay people, and pay our respects to elders past and present.

This research has made use of the SIMBAD database, operated at CDS, Strasbourg, France \citep{Simbad}.
This research has made use of NASA’s Astrophysics Data System Bibliographic Services.

This paper made use of the Whole Sky Database (wsdb) created by Sergey Koposov and maintained at the Institute of Astronomy, Cambridge by Sergey Koposov, Vasily Belokurov and Wyn Evans with financial support from the Science \& Technology Facilities Council (STFC) and the European Research Council (ERC).

This work has made use of data from the European Space Agency (ESA) mission
{\it Gaia} (\url{https://www.cosmos.esa.int/gaia}), processed by the {\it Gaia}
Data Processing and Analysis Consortium (DPAC,
\url{https://www.cosmos.esa.int/web/gaia/dpac/consortium}). Funding for the DPAC
has been provided by national institutions, in particular the institutions
participating in the {\it Gaia} Multilateral Agreement.

This project used public archival data from the Dark Energy Survey
(DES). Funding for the DES Projects has been provided by the
U.S. Department of Energy, the U.S. National Science Foundation, the
Ministry of Science and Education of Spain, the Science and Technology
Facilities Council of the United Kingdom, the Higher Education Funding
Council for England, the National Center for Supercomputing
Applications at the University of Illinois at Urbana-Champaign, the
Kavli Institute of Cosmological Physics at the University of Chicago,
the Center for Cosmology and Astro-Particle Physics at the Ohio State
University, the Mitchell Institute for Fundamental Physics and
Astronomy at Texas A\&M University, Financiadora de Estudos e
Projetos, Funda{\c c}{\~a}o Carlos Chagas Filho de Amparo {\`a}
Pesquisa do Estado do Rio de Janeiro, Conselho Nacional de
Desenvolvimento Cient{\'i}fico e Tecnol{\'o}gico and the
Minist{\'e}rio da Ci{\^e}ncia, Tecnologia e Inova{\c c}{\~a}o, the
Deutsche Forschungsgemeinschaft, and the Collaborating Institutions in
the Dark Energy Survey.  The Collaborating Institutions are Argonne
National Laboratory, the University of California at Santa Cruz, the
University of Cambridge, Centro de Investigaciones Energ{\'e}ticas,
Medioambientales y Tecnol{\'o}gicas-Madrid, the University of Chicago,
University College London, the DES-Brazil Consortium, the University
of Edinburgh, the Eidgen{\"o}ssische Technische Hochschule (ETH)
Z{\"u}rich, Fermi National Accelerator Laboratory, the University of
Illinois at Urbana-Champaign, the Institut de Ci{\`e}ncies de l'Espai
(IEEC/CSIC), the Institut de F{\'i}sica d'Altes Energies, Lawrence
Berkeley National Laboratory, the Ludwig-Maximilians Universit{\"a}t
M{\"u}nchen and the associated Excellence Cluster Universe, the
University of Michigan, the National Optical Astronomy Observatory,
the University of Nottingham, The Ohio State University, the OzDES
Membership Consortium, the University of Pennsylvania, the University
of Portsmouth, SLAC National Accelerator Laboratory, Stanford
University, the University of Sussex, and Texas A\&M University.
Based in part on observations at Cerro Tololo Inter-American
Observatory, National Optical Astronomy Observatory, which is operated
by the Association of Universities for Research in Astronomy (AURA)
under a cooperative agreement with the National Science Foundation.


{\it Facilities:} 
{Anglo-Australian Telescope (AAOmega+2dF)}

{\it Software:} 
{\code{numpy} \citep{numpy}, 
\code{scipy} \citep{scipy},
\code{matplotlib} \citep{matplotlib}, 
\code{seaborn} \citep{seaborn},
\code{astropy} \citep{astropy,astropy:2018},
\code{RVSpecFit} \citep{rvspecfit}
\code{q3c} \citep{Koposov2006}, 
\code{emcee} \citep{emcee},
\code{gala} \citep{gala, adrian_price_whelan_2020_4159870},
\code{galpot} \citep{Dehnen:1998}
}

\newpage

\bibliography{main}{}
\bibliographystyle{aasjournal}



\end{document}